\documentclass[a4paper,11pt]{article}
\pdfoutput=1

\usepackage{jheppub}
\usepackage[T1]{fontenc}
\usepackage{tensor}
\usepackage{subcaption}
\hypersetup{bookmarksnumbered}
\usepackage{bbold}
\usepackage{float}
\usepackage{mathtools}
\usepackage[table]{xcolor}
\usepackage{longtable}
\usepackage[labelformat=simple]{subcaption}

\usepackage[colorinlistoftodos]{todonotes}

\usepackage[smalltableaux]{ytableau}

\DeclareFontFamily{OT1}{pzc}{}
\DeclareFontShape{OT1}{pzc}{m}{it}{<-> s * [1.10] pzcmi7t}{}
\DeclareMathAlphabet{\mathpzc}{OT1}{pzc}{m}{it}

\newcommand{\bZ}{\mathbb Z}

\def\legP{\mathcal{P}}

\def\avg#1{\left<#1\right>}
\def\tr#1{\text{Tr}\left(#1\right)}

\title{\boldmath Where is tree-level string theory?}

\author[a,b]{Jan Albert,}
\author[a]{Waltraut Knop}
\author[a]{and Leonardo Rastelli}

\affiliation[a]{C. N. Yang Institute for Theoretical Physics, Stony Brook University,\newline Stony Brook, NY 11794-3840, U.S.A.}

\affiliation[b]{Simons Center for Geometry and Physics, Stony Brook University,\newline Stony Brook, NY 11794-3636, U.S.A.}

\preprint{YITP-SB-2024-12}

\abstract{
We investigate the space of consistent tree-level extensions of the maximal supergravities in ten dimensions.
We parametrize theory space by the first few EFT coefficients
and by the on-shell coupling of
 the lightest massive state, and impose on these data
 the constraints
that follow from $2 \to 2$ supergraviton scattering.  While Type II string theory lives strictly inside the  allowed region, we uncover a novel extremal solution of the bootstrap problem, 
which appears to contain a {\it single} linear Regge trajectory, with the 
same slope as  string theory. 
We repeat a similar analysis for supergluon scattering, where we find instead a continuous family of extremal solutions
with a single Regge trajectory
of varying slope.
}

\begin{document} 
\maketitle

\section{Introduction and summary}

Gravity is universal at large distance, 
where it is described by classical general relativity. At short distance the gravitational coupling grows, and Einstein's theory must be UV completed. Determining the space of consistent UV completions of Einstein gravity is 
a theoretical question of fundamental importance. 
In asymptotically flat space, consistency 
should be understood in terms of a relatively well-established set of axiomatic requirements
for the gravitational S-matrix, such as unitarity and causality. 

We will be concerned with modifications of Einstein gravity that occur already at {\it tree level}.
String theory provides the prototypical example.
 New degrees of freedom (massive higher-spin states) appear at an energy scale $M_{\rm string}$,
which at weak coupling $g_s \ll 1$ is parametrically smaller
than the Planck scale, $M_{\rm string} \ll  M_{\rm Planck}$.  Tree-level string amplitudes
are of course causal  and unitary, but in ways that appear absolutely miraculous if one does not refer to the underlying worldsheet picture. One
may legitimately wonder whether string theory is in fact the unique framework to modify gravity at tree level. For the question to make sense, one needs to spell out precisely the consistency conditions to be imposed, as we shall do. Various conjectures
in this direction have been 
formulated in~\cite{Chowdhury:2019kaq, Chakraborty:2020rxf}. There is also a line of work that aims to probe the rigidity of  string theory by looking for explicit deformations of tree-level string amplitudes that obey at least some of the desired properties~\cite{Cheung:2022mkw, Cheung:2023adk, Rigatos:2023asb, Cheung:2023uwn, Arkani-Hamed:2023jwn, Cheung:2024uhn, Bhardwaj:2024klc}.

This question can be systematically addressed within the framework of the modern S-matrix
bootstrap (see~e.g.~\cite{Kruczenski:2022lot, deRham:2022hpx} for reviews).
 A very useful window into the problem is the study of constraints on effective field theory (EFT)~\cite{Adams:2006sv, Arkani-Hamed:2020blm, Bellazzini:2020cot, Tolley:2020gtv, Caron-Huot:2020cmc}.
Rigorous  two-sided bounds on Wilson coefficients of the higher-derivative terms 
(normalized by the Newton constant and in units of the EFT cut-off) have been obtained in~\cite{Caron-Huot:2021rmr,Caron-Huot:2022ugt, Caron-Huot:2022jli} by exploiting the consistency conditions encoded in $2 \to 2$ graviton scattering. Here we continue this research program, with an eye towards interpreting the EFT bounds in terms of actual amplitudes that saturate them, in the spirit of~\cite{Caron-Huot:2020cmc}. 

The cleanest arena for this kind of investigations are the theories with maximal supersymmetry. In this paper 
we will constrain the possible tree-level\footnote{Our exploration of the space of {\it tree-level} extensions of supergravity should be contrasted with the work of~\cite{Guerrieri:2021ivu, Guerrieri:2022sod},
%(whose titles we echo), 
which considered a similar question but for the full quantum amplitudes in string theory and  M-theory.}  completions of
the {\it ten-dimensional maximal supergravities}, continuing the analysis initiated in~\cite{Caron-Huot:2021rmr}. Technically, maximal supersymmetry offers some drastic simplifications. Physically, few would probably doubt that Type IIA 
string theory provides the 
 only fully  consistent tree-level extension of non-chiral 10D maximal supergravity, and similarly that Type IIB gives  the only such extension in the chiral case.
Establishing these conjectures by abstract bootstrap methods is plausibly within the reach of existing technology, and  would be an important benchmark.

\subsubsection*{Assumptions}
Our assumptions are rather
minimal, and straightforward to state:
\begin{enumerate}
\item[(i)] {\it Spectrum}.  The only massless particles are the graviton and its superpartners.  We also  require that there is a gap to the next massive state, in other terms we forbid an accumulation point  at zero mass. Furthermore, the spin of  single-particle states
of a given mass should be bounded from above -- this can be understood as a locality requirement.
\item[(ii)] {\it Analyticity.}  $2\to 2$ amplitudes are analytic anywhere
except for singularities on the real axis in the complex $s$ plane (where $s$ is the usual Mandelstam variable). These are the physical singularities due to tree-level
exchanges.
\item[(iii)]  {\it Crossing symmetry}.
\item[(iv)]  {\it Unitarity.} At tree level, unitarity reduces to (semidefinite) positivity of the partial wave expansion.
E.g., in the simplest case of a $2\to 2$ amplitude of identical  scalars,  the coefficients of the partial wave expansion must be non-negative.
\item[(v)] {\it Strict spin-two Regge boundedness}.
 $2 \to 2$ amplitudes grow {\it strictly slower} than $O(s^2)$ in the Regge  limit of large $s$
and fixed   
physical momentum transfer.
\end{enumerate}
The only assumption that requires some elaboration is the last one. The slightly weaker requirement that tree-level amplitudes should grow at most like  $O(s^2)$ in the Regge limit
is, we believe, uncontroversial.\footnote{This is the ``classical Regge growth conjecture'' of~\cite{Chowdhury:2019kaq}, supported by a variety  of compelling arguments, notably the analysis of CEMZ~\cite{Camanho:2014apa}
and especially the connection with the ``chaos bound'' in holographic theories, as discussed
in~\cite{Chandorkar:2021viw}.} Pure classical supergravity, of course, grows exactly as $O(s^2)$, because of the graviton exchange, but here
we are insisting that its tree-level extensions 
should have {\it strictly slower} Regge growth. 
Because of maximal supersymmetry, any massive state belongs to a multiplet that reaches {\it at least} up to spin $J=4$. As shown by CEMZ~\cite{Camanho:2014apa}, any tree-level exchange of spin $J >2$ induces a causality violation that can only be compensated by an  infinite tower of states of arbitrarily high spin. On very general grounds, these states must arrange in Regge trajectories.
If we only granted $O(s^2)$ Regge behavior and not better, analyticity in spin would a priori be guaranteed only for $J>2$ and the graviton might be an isolated state that does not belong to any trajectory. We are insisting instead that
the leading trajectory extends down to
the graviton, so that its intercept $\alpha (u) = 2 + \alpha' u/4  + O(u^2)< 2$ for small 
momentum transfer $u$ in the physical region $u<0$. 
In other terms, our assumption (v) is equivalent to the statement that the graviton should reggeize.\footnote{There are  also rigorous estimates~\cite{Haring:2022cyf} showing that  the  full quantum gravitational amplitude must have Regge behavior strictly better than $O(s^2)$. 
In~\cite{Haring:2022cyf}, the Regge behavior at
{\it tree-level} (for theories with a perturbative expansion) was argued to be controlled  by the ``local growth'' of the quantum amplitude (fixed ${\rm Re} \,s$ for ${\rm Im} \,s \to \infty$) 
in the weak-coupling limit, and shown to be at least as good as $O(s^2)$. It would be very interesting to see whether this conclusion can be slightly sharpened, if one assumes a hierarchy of scales. We envision a continuous family of amplitudes, characterized by an additional EFT scale $M$ in addition to the Planck scale $M_{\rm Planck}$, where $M/M_{\rm Planck}$ is a parameter that can take an arbitrarily small value, so that the theory is weakly coupled for
 $M/M_{\rm Planck} \to 0$.  (Pure gravity does not conform to this scenario because it has a single scale, $M_{\rm Planck}$.) 
} 

A more pedestrian justification of  (v) 
is the desire to avoid ``trivial'' continuous families of solutions to the bootstrap problem. 
If we only assume $O(s^2)$ Regge behavior, pure supergravity is a consistent solution. As convex linear combinations of solutions are solutions, starting from any 
consistent solution we can 
find a one-parameter family of them by just rescaling at will the value of the Newton constant. 
Indeed, the value of the Newton constant does not enter any the bootstrap equations, because no dispersion relation can constrain it if the allowed Regge growth is $O(s^2)$. 
In this sense, (v) can be regarded as part of 
any reasonable definition of what one can mean by a ``tree-level UV completion of gravity'': without it, gravity just decouples from the bootstrap problem.

\subsubsection*{Setup}

Our general setup is the same as in~\cite{Caron-Huot:2021rmr}.
The simplest set of constraints, and the only ones that we will study in this paper, come from the $2 \to 2$ amplitude of supergravitons. In 
both the chiral and nonchiral 10D maximal supergravity theories, thanks to supersymmetric Ward identities
the full superamplitude can be obtained from an auxiliary fully crossing symmetric {\it scalar} amplitude $M(s, u)$, a key technical simplification. What's more, the $J <2$ Regge  behavior 
of the graviton amplitude translates into $J< -2$ for the auxiliary amplitude,
\begin{equation} \label{minustwo}
\lim_{|s| \to \infty} s^2 M(s, u) \to 0 \quad {\rm for \;fixed}\; {u<0}\,.
\end{equation}
 In a low-energy expansion,
 \begin{equation}
M(s, u) = \frac{8 \pi G}{stu} + g_0 + g_2 (s^2 + t^2 + u^2) + \dots\,.
\end{equation}
The supergraviton exchange corresponds to the first 
 term, proportional to the Newton constant $G$.
Its Regge behavior is exactly minus two, violating (\ref{minustwo}).
  Any correction  signals 
  a deviation from pure supergravity and
  the presence of higher massive states.
For example, the $g_0$ Wilson coefficient is related to the  $R^4$ term in the effective action, the first higher-derivative correction allowed by maximal supersymmetry. 
There is by a now a well-established technology~\cite{Arkani-Hamed:2020blm, Bellazzini:2020cot, Tolley:2020gtv, Caron-Huot:2020cmc} to carve out the space of EFTs consistent with causality and positivity. Dispersion relations relate the IR to the UV, yielding sum rules for the
the low-energy Wilson coefficients in terms of an unknown high-energy spectral density, as well as ``null constraints''~\cite{Tolley:2020gtv, Caron-Huot:2020cmc} that encode crossing symmetry.
Positivity of the spectral density allows to set up a semidefinite program that yields rigorous bounds on the Wilson coefficients.
Meaningful bounds are possible 
for their homogeneous ratios in units of cut-off $M$, for example $\tilde g_0 = g_0 M^6/(8 \pi G)$ and $\tilde g_2 = g_2 M^{10}/(8 \pi G)$. Because of the graviton pole $\sim 1/u$,
one needs to face the technical complication that
not all sum rules  can be expanded in the forward limit $u \to 0$. The sum rule for the Newton constant
is dealt with~\cite{Caron-Huot:2021rmr} by {\it smearing}\footnote{For a variety of perspectives about the interplay between positivity bounds and the graviton pole, see~e.g.~\cite{Alberte:2020jsk,Tokuda:2020mlf, Herrero-Valea:2020wxz, Alberte:2021dnj, Herrero-Valea:2022lfd, Noumi:2022wwf, Caron-Huot:2024tsk}.} with a suitable set of test functions of $u$. 

It is also possible to integrate-in massive states, viewing the first (or first few) of them as part of the low-energy EFT. Their normalized on-shell couplings become part of the data that can be bounded.

\subsubsection*{Results}

The preliminary exploration of~\cite{Caron-Huot:2021rmr} derived an upper bound on $\tilde g_0$,
which is obeyed (as it must)  by tree-level Type II string theory,
but unfortunately not saturated by it. 
Here we carry out a more systematic investigation. 

The simplest way to refine the analysis  is to derive exclusion plots in the space of multiple EFT couplings, 
such as~$\tilde g_0$ and $\tilde g_2$,
as in figure \ref{fig:CSg0g2}. We  again find that string theory lives safely within the bounds. 
The exclusion boundary has two interesting kinks. 
At this stage, we are able to identify one of them.
It
corresponds to an unphysical amplitude that contains  
the graviton and  infinitely many states  with the same mass $m = M$ (i.e.~right at the cut-off) of arbitrarily high spin. This violates one of our spectral assumptions in (i). We'll come back to the second kink momentarily.

To enforce locality, we insist 
 for only a {\it finite} number of spins at the lowest non-zero mass $m$, with a new cut-off $M >m$. Our numerics indicate that the only consistent possibility is to have a {single} state $\phi$ at $m= m_\phi$, with ``effective spin'' zero. By effective spin we mean the spin of an exchanged particle in the auxiliary scalar amplitude $M(s, u)$. In terms of the physical superamplitude, this corresponds to a whole massive multiplet whose highest spin is $J_{\rm max} = 4$ (in general,  $J_{\rm max}= J_{\rm eff} + 4$).

\begin{figure}[ht]
\centering
\includegraphics[width=0.8\textwidth]{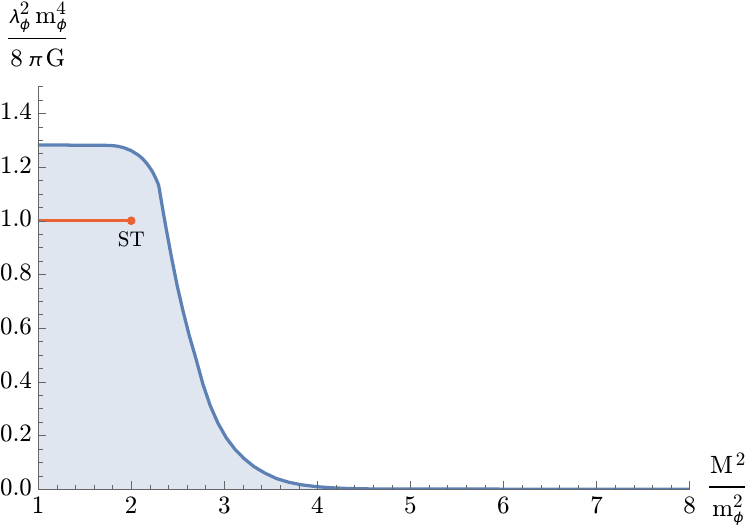}
\caption{Exclusion plot for the normalized on-shell coupling
of an isolated (effective) scalar of mass $m_\phi$, as a function of the cut-off $M$. Allowed values are in the blue region.}
\label{fig:CSgPM1noNmax}
\end{figure}

We perform a thorough analysis of the setup where 
the effective scalar\footnote{From now on, we will usually drop  the qualifier ``effective''. Unless otherwise stated, it is  understood that we are thinking in terms of the auxiliary scalar amplitude.} of mass $m_\phi$ has been integrated-in.
 Perhaps the most illuminating plot is 
figure~\ref{fig:CSgPM1noNmax} (a streamlined version of figure \ref{fig:CSgPM1}), where we maximize 
the on-shell coupling of the scalar as a function of the cut-off.
The numerical exclusion region has a flat plateau that drops sharply
at $M \cong \sqrt{2} m_\phi$.
Remarkably, in string theory we also have $M^2 =2 m_\phi^2$
(this is the standard relation between the masses of the first and second levels), but the string on-shell coupling 
(red line in figure~\ref{fig:CSgPM1noNmax})
is strictly smaller than the value at our plateau.

We interpret the endpoint of the plateau  in terms of a very intriguing extremal solution of the boostrap problem, 
which also explains the {\it other} kink in figure~\ref{fig:CSg0g2}.
Directly extracting its spectrum from the numerical semidefinite solver is a bit subtle, 
but by a variety of experiments involving different spectral assumptions we conclude that the extremal solution contains a {\it single}, linear Regge trajectory.
This is reminiscent of recent findings for pion scattering~\cite{Albert:2023seb}, where forcing the presence of a spin two resonance (and maximizing its coupling) leads to an extremal amplitude with a single (but curved in that case) Regge trajectory. The intuition is similar. A  single massive scalar exchange has Regge growth zero, which violates our Regge assumption by a finite amount. The only way to fix this is to have (at least) a whole Regge trajectory, as anticipated by CEMZ~\cite{Camanho:2014apa}. 
The upshot is that
we assumed {\it almost nothing} and got a linear Regge trajectory
with the same slope as in string theory. 
That the graviton should reggeize is expected from our assumption (v), but it is certainly non-trivial to see this feature emerge naturally at a special point of the exclusion boundary.

The conceptual status of this curious  extremal solution with a single linear Regge trajectory is not fully clear to us. Under rather general assumptions, one can rigorously rule out the existence of  amplitudes with a single trajectory~\cite{Eckner:2024pqt}. Perhaps 
our numerical solution is only approximate (e.g.~new states come in at very  high mass). Or there may exist an exact solution with a single trajectory
that violates one of the assumptions in~\cite{Eckner:2024pqt}; the most likely scenario would be the presence of Regge cuts. In the latter case, it would be very interesting to find an explicit analytic form, as we have tried but not yet succeeded to do.\footnote{Curiously, the primal approach of~\cite{Guerrieri:2022sod} uncovered extremal solutions with daughter trajectories. Of course, there is no contradiction, as we are assuming tree level while they are working in the nonperturbative regime. It would be valuable to develop a conceptual understanding of why the two bootstrap approaches give qualitatively different extremal solutions.}

We still have not cornered string theory, which has, of course, infinitely many daughter trajectories. But we should be optimistic. So far we have only used the constraints from $2\to 2$
supergraviton scattering. A  natural next step is to study the full mixed system of $2 \to 2$ amplitudes where
the supergraviton and the first massive multiplet are allowed as external legs~\cite{inprogress}. It is not too far fetched to speculate that string theory may be singled out as the unique solution of these  mixed system constraints.

\subsubsection*{Extension to open strings}

We can repeat a similar analysis for
the scattering of 10D supergluons, with 
the goal of cornering the open string sector of Type I string theory. While for supergravity we were able to setup the problem
by making a minimal, universal set of assumptions that should  apply to any general tree-level completion, 
here the story is  somewhat less canonical. We need to input some features that from the start   put us broadly in the realm of open string amplitudes,
such as the Chan-Paton structure of colored-ordered amplitudes
(which implies that all exchanged particles are in the adjoint representation), and Regge behavior $J<1$ from reggeization of the gluon. Thanks to $(1, 0)$ 
supersymmetry, the full superamplitude is 
again captured by a single auxiliary scalar amplitude $M(s, u)$, which is now
only $s \leftrightarrow u$ crossing symmetric, and with Regge behavior $J<-1$. The numerical exploration is parallel to the gravitational case, but the results
quite different. 
Integrating-in an isolated   massive ``scalar'' (corresponding to a massive supermultiplet with spin up to two in the physical superamplitude) and maximizing its coupling leads to
figure~\ref{fig:OSgPM1}, the open string analog of 
 figure~\ref{fig:CSgPM1noNmax}. 
 Type I string theory lives strictly inside the exclusion region.
 Rather than an approximate step function, the exclusion boundary is now smoothly curved. We find that each point of the curved boundary corresponds to an extremal amplitude with a single linear Regge trajectory, whose slope varies as we move along the boundary. We find no apparent special features for the string theory value of the slope.

\bigskip
\noindent
The detailed organization of the paper is best
apprehended from the table of contents. In section~\ref{sec:Ward}
we review supersymmetric Ward identities.
In sections~\ref{sec:closedStringsSetup} and~\ref{sec:CSResults}
we describe the general setup and present our numerical results
for supergraviton scattering, respectively. Similarly, sections~\ref{sec:openStrings} and~\ref{sec:OSResults} deal with the setup and results for open strings. We conclude in section~\ref{sec:Outlook}
with a brief outlook. Appendix \ref{app:numerics} contains some technical details of the numerical implementation.

\bigskip
\noindent {\bf Note added:} As we were finalizing this work, we received the very nice paper~\cite{Berman:2024wyt}, which performs an analysis broadly similar to ours for the case of open strings. The general setup is the same.
Our assumption of $J<1$ Regge behavior and the  implementation of the smearing procedure allow us to normalize by the Yang-Mills coupling $g^2_{\rm YM}$, while \cite{Berman:2024wyt}
considers ratios of higher Wilson coefficients. 
Our results are largely complementary, and appear to be entirely compatible with theirs.

\section{Supersymmetric Ward identities}

\label{sec:Ward}

In this work we are concerned with the $2\to 2$ scattering of massless supermultiplets in $D=10$ with various amounts of supersymmetry. For $\mathcal N = (2,0)$ and $\mathcal N = (1,1)$ we will consider the scattering of the graviton supermultiplet, and for $\mathcal N = (1,0)$ the scattering of the super Yang-Mills multiplet. The three cases share the simplifying feature that all four-point amplitudes between components of the supermultiplet are encoded in terms of a single auxiliary scalar amplitude. We start by briefly reviewing case by case how this follows from supersymmetry Ward identities %\cite{Grisaru:1976vm,Grisaru:1977px}
(see~e.g.~\cite{Elvang:2013cua} for a pedagogical introduction). Readers who are already familiar with this fact might want to proceed directly to section~\ref{sec:closedStringsSetup}.

\subsection{$\mathcal N = (2,0)$}
Consider first the case $\mathcal N=(2,0)$ with two (real) supercharges $Q_1^A,Q^A_2$ of the same chirality. Apart from the graviton, the massless supermultiplet contains the dilaton, the Kalb-Ramond two-form $B$-field, RR forms $C^{(i)}$ of rank zero,  two and four (the latter being selfdual), and two gravitinos (as well as dilatinos) of the same chirality. To construct it, we start from the SUSY algebra in the massless representation,
\begin{equation}
    \left\{Q_i^A,Q_j^B\right\} = \lambda^{A,a}\lambda^{B}_a\delta_{ij}\,,
\end{equation}
where $\lambda^{A,a}$ are $D=10$ spinor-helicity variables \cite{Caron-Huot:2010nes,Boels:2012ie} satisfying
\begin{equation}
    p_\mu \sigma^{\mu,BA'} = \lambda^{B,a}\lambda_a^{A'}\,, \qquad p^2=0\,.
\end{equation}
We use conventions where $A,A'=1,...,16$ denote Weyl indices of opposite chirality and lowercase indices $a,a'=1,...,8$ indicate spinor and conjugate spinor indices of the $SO(8)$ little group.\footnote{The latter can be raised and lowered with $\delta_{ab}$, $\delta_{a'b'}$ and their inverses. Instead of raised/lowered, spacetime Weyl indices $A,A'$ can be exchanged using the components of the charge conjugation matrix
$$
C= \begin{pmatrix}
        0 & \Omega\indices{_B^{A'}} \\
        \Omega\indices{^{B'}_A} & 0
    \end{pmatrix}\,.
$$
}

On this massless representation, half of the supercharges are inactive \cite{Strathdee:1986jr} and we can use the expansion
\begin{equation}
    Q_1^A = q_{1a} \lambda^{A,a}\,, \qquad Q_2^A = q_{2a} \lambda^{A,a}\,,
\end{equation}
with $q_{ia}$ parametrizing the \textit{active} supercharges. We can in turn arrange these supercharges into (complex) linear combinations $q^{\pm}_a$ with definite $SO(2)_R$ $R$-charge $\pm1$ satisfying
\begin{equation}
    \left\{q^{+}_a,q^{+}_b\right\}= \left\{q^-_a,q^-_b\right\} = 0\,, \qquad
    \left\{q^{+}_a,q^-_b\right\} = \delta_{ab}\,.
\end{equation}
The massless multiplet is then constructed by repeatedly acting with the raising operator $q^{+}_a$ on a (scalar) Clifford vacuum of $R$-charge $-4$. The top and bottom of the multiplet constructed in this way are thus scalars of $R$-charge $\pm 4$ and correspond to different linear combinations of the dilaton and the axion $C^{(0)}$.

The $2\to 2$ scattering amplitudes of all the components in the multiplet are compactly encoded in a superamplitude $\mathcal A_4(p_i,\eta_i)$, depending on the momenta $p_i$ and fermionic coordinates $\eta_i$ parametrizing an on-shell superspace for each external leg. One should think of $\mathcal A_4$ as a function of four superfields of the form\footnote{We note that the on-shell supersymmetry algebra is realized on this superspace by the choice
\begin{equation}
    q^{+}_a = \eta_a\,, \qquad q^{-}_a = \frac{\partial}{\partial \eta^a}\,,
\end{equation}
since $\{\eta_a,\tfrac{\partial}{\partial\eta^b}\}=\delta_{ab}$.}
\begin{equation}
    \Phi(p,\eta) = \varphi(p) + \eta^a\psi_a(p) + \frac{1}{2!}\eta^a\eta^b \varphi_{ab}(p) + \cdots + \eta^1\eta^2...\eta^8 \bar\varphi(p) \,,
\end{equation}
where the terms $\varphi_{a,b,...}(p),\psi_{a,b,...}(p)$ describe respectively bosonic and fermionic components of the supermultiplet with momentum $p$ and little group indices $a,b,...\,$.
One can then extract from $\mathcal A_4$ any $2\to 2$ amplitude involving different components of the supermultiplet by taking suitable derivatives $\frac{\partial}{\partial \eta^a_i}$ for each external leg. We will be particularly interested in the scattering of the complex scalar $\varphi(p)$ --- the axi-dilaton.

Supersymmetry imposes nontrivial constraints on scattering amplitudes in the form of the so-called SUSY Ward identities \cite{Grisaru:1976vm,Grisaru:1977px}. For $\mathcal N=(2,0)$ supergraviton scattering, these were solved in \cite{Boels:2012ie} ---here we simply review their results.
In terms of the superamplitude $\mathcal A_4$, the SUSY Ward identities read
\begin{equation}\label{eq:SuperWard}
    Q_+^{A'} \mathcal A_4 = 0\,, \qquad Q_{-}^{A} \mathcal A_4 = 0\,,
\end{equation}
where $Q_{\pm}$ are the ``total'' supercharges defined by the following sums over external legs:
\begin{equation}
    Q_+^{A'} \equiv \sum_{i=1}^4 i^{A'}_a \eta_i^a\,, \qquad Q_-^A \equiv \sum_{i=1}^4 i^{A,a}\frac{\partial}{\partial \eta_i^a}\,,
\end{equation}
where $i^{A,a}$ refers to the spinor-helicity variable $\lambda^{A,a}$ for the external particle $i$.

In the current case of four-particle scattering, there is a unique solution to the SUSY Ward identities \eqref{eq:SuperWard}. It can be written in an explicitly Lorentz-invariant way using the ``fermionic delta function''
\begin{equation}
    \delta^{16}(Q_+) \equiv \frac{1}{16!} \epsilon_{A_1'...A_{16}'}Q_+^{A_1'}\cdots Q_+^{A_{16}'}\,.
\end{equation}
Making the momentum-conservation delta function explicit, the solution is
\begin{equation}\label{eq:A4}
    \mathcal A_4(p_i,\eta_i) = \delta^{10}(p_1+p_2+p_3+p_4)\delta^{16}(Q_+) M(s,u)\,,
\end{equation}
with $M(s,u)$ a bosonic function of the Mandelstam invariants
\begin{equation}
    s = - (p_1 + p_2)^2 \, , \quad t = - (p_2 + p_3)^2\, , \quad u = - (p_1 + p_3)^2 \,.
\end{equation}
It is clearly killed by $Q_+^{A'}$ by antisymmetry. Upon acting by $Q_-^B$, the supersymmetry algebra guarantees that we get an expression proportional to the overall momentum
\begin{equation*}
    \left\{ Q_-^{B}, Q_+^{A'}\right\} = \sum_{i=1}^4 i^{B,a} i^{A'}_a = \left(\sum_{i=1}^4 {p_i}_\mu\right) \sigma^{\mu,BA'}\,,
\end{equation*}
which then vanishes by momentum conservation.

That there are no other solutions follows from $SO(2)_R$ $R$-charge counting. Each fermionic coordinate $\eta$ contributes $R$-charge $+1$, so the fermionic delta function $\delta^{16}(Q_+)$ has $R$-charge $+16$. This exactly balances the contribution from the Clifford vacua in four-point amplitudes. For higher-point amplitudes one would need to add a function $f(\eta_i)$ with higher powers of the fermionic coordinates to balance the $R$-charge deficit. This would potentially lead to multiple solutions of the SUSY Ward identities. But, for four-point amplitudes, $f(\eta_i)=1$ and the solution presented above is unique.

The important lesson from \eqref{eq:A4} is that all $2\to 2$ scattering amplitudes involving any components of the graviton supermultiplet are determined in terms of a \textit{single} bosonic amplitude $M(s,u)$. For example, we can extract the axi-dilaton self-scattering amplitude by looking at the term\footnote{Here we used the identity
\begin{equation}
    \epsilon_{A'_1\ldots A'_{16}}3^{A'_1}_{a_1}...3^{A'_8}_{a_8}4^{A'_9}_{b_1}...4^{A'_{16}}_{b_8}
    = \epsilon_{a_1...a_8}\epsilon_{b_1...b_8}(2 p_3\cdot p_4)^4\,,
\end{equation}
see \cite{Boels:2012ie} for other useful identities of a similar form.
}
\begin{align}
    \mathcal A_4 \supset&\, \begin{pmatrix}
        16 \\ 8
    \end{pmatrix}
    \frac{1}{16!}\epsilon_{A'_1\ldots A'_{16}}3^{A'_1}_{a_1}...3^{A'_8}_{a_8}4^{A'_9}_{b_1}...4^{A'_{16}}_{b_8}\eta_3^{a_1}...\eta_3^{a_8}\eta_4^{b_1}...\eta_4^{b_8}  \delta^{10}({\textstyle \sum_i p_i}) M(s,u) \nonumber\\
    =&\, \left(\frac{1}{8!} \epsilon_{a_1...a_8}\eta_3^{a_1}...\eta_3^{a_8}\right)
    \left(\frac{1}{8!}\epsilon_{b_1...b_8}
    \eta_4^{b_1}...\eta_4^{b_8}\right) \delta^{8}({\textstyle \sum_i p_i}) s^4 M(s,u)\,,
\end{align}
from which we extract
\begin{equation}\label{eq:axi-dilaton}
    A\left(\varphi\varphi\bar\varphi\bar\varphi\right) = s^4 M(s,u)\,.
\end{equation}
Since it involves complex scalars, this amplitude is a priori only crossing symmetric under $1\leftrightarrow 2$ or $3\leftrightarrow 4$ exchanges. But now we see that supersymmetry enhances its crossing properties. Indeed, the superamplitude \eqref{eq:A4} involves four identical superfields and so it must be fully crossing symmetric. This forces $M(s,u)$ (and therefore $\frac{1}{s^4} A(\varphi\varphi\bar\varphi\bar\varphi)$) to be fully crossing-symmetric under $s\leftrightarrow t \leftrightarrow u$.
\subsection{$\mathcal N = (1,0)$}\label{sec:N=(1,0)}
Let us now turn to the less supersymmetric case of $\mathcal N = (1,0)$, with a single (real) supercharge $Q^A$. In this case there are two massless representations: the graviton and the SYM multiplets. We will focus on the latter. This multiplet contains the $SO(8)$ little group representations $(8_v + 8_c)$, describing gauge fields and gauginos. On this massless representation, we can again write $Q_1^A = q_{a} \lambda^{A,a}$ with
\begin{equation}
    \left\{q_a,q_b\right\} = \delta_{ab}\,.
\end{equation}
The difference with the previous case is that, since all eight real supercharges are in an irreducible representation $q_{a}$, we cannot arrange them into raising and lowering operators while keeping explicit $SO(8)$ covariance. There is therefore no immediate generalization of the superspace used above for this case. However, since we will only be concerned with four-point amplitudes ---and these can always be embedded in a four-dimensional space by suitable Lorentz transformations--- we will be able to use the more familiar four-dimensional SUSY language, for which an on-shell superspace is known \cite{Elvang:2009wd,Elvang:2010xn}.

When restricting to four dimensions, the 16 (real) supercharges rearrange into the complex Weyl spinors $Q_\alpha^I, \bar Q_{J\dot \beta}$ $(I,J=1,...,4)$, generating $\mathcal N=4$ supersymmetry. On the massless multiplet they satisfy the algebra
\begin{equation}
    \left\{Q_\alpha^I, \bar Q_{J\dot \beta}\right\} = |p]_{\alpha} \langle p|_{\dot \beta}\, \delta^I_J,
\end{equation}
where $|p]$, $\langle p|$ are the usual spinor-helicity variables, $\alpha,\dot\beta$ denote four-dimensional Weyl indices of opposite chirality, and upper/lower $I,J$ indices denote fundamental/antifundamental $SU(4)_R$ $R$-symmetry indices. An on-shell superspace for this representation is obtained by introducing four fermionic coordinates $\eta^I$ and setting \cite{Elvang:2009wd,Elvang:2010xn}
\begin{equation}
    Q_\alpha^I = |p]_{\alpha} \frac{\partial}{\partial \eta_I}\,, \qquad \bar Q_{J\dot \beta} = \langle p|_{\dot \beta} \eta_J\,.
\end{equation}

The SYM multiplet is obtained by repeatedly acting by the different $\bar Q$ on a Clifford vacuum with helicity $-1$, and it can be encoded in the following superfield:
\begin{equation}
    \Phi(p,\eta) = \varphi_- + \eta_I \psi^I + \frac{1}{2!}\eta_I\eta_J\varphi^{IJ} + \frac{1}{3!} \eta_I\eta_J\eta_K \psi^{IJK} + \eta_1\eta_2\eta_3\eta_4 \varphi_+\,.
\end{equation}
From a four-dimensional point of view this is the usual field content of $\mathcal N=4$ SYM; a gauge field $\varphi_+,\varphi_-$, six scalars $\varphi^{IJ}$ and four fermions of each chirality $\psi^I, \psi^{IJK}$. From a ten-dimensional point of view, the six scalars provide the missing polarizations to the gauge field, combining into the $8_v$. The fermions similarly combine into the $8_c$. With this superspace at hand it is immediate to write down a superamplitude $\mathcal A_4(p_i,\eta_{i})$ solving the SUSY Ward identities \cite{Elvang:2009wd,Elvang:2010xn}. Namely,
\begin{equation}\label{eq:A4-(1,0)}
    \mathcal A_4(p_i,\eta_i) = \delta^{10}(p_1+p_2+p_3+p_4)\delta^{8}(\bar Q) M(s,u)\,,
\end{equation}
where now the fermionic delta function is given by
\begin{equation}
    \delta^{8}(\bar Q) \equiv \frac{1}{2^4} \prod_{I=1}^4 \bar Q_{I\dot \beta} \bar Q_{I}^{\dot \beta} = \frac{1}{2^4} \prod_{I=1}^4 \sum_{i,j=1}^4 \langle ij\rangle  \eta_{iI}\eta_{jJ}\,.
\end{equation}
Like in the case above, the solution is killed by the total supercharge $\bar Q$ by antisymmetry and by $Q$ because its commutator is proportional to the total momentum. Up to the bosonic amplitude $M(s,u)$, this solution is unique, as can be seen again from $R$-symmetry counting: The $U(1)_R$ charge $+8$ of the fermionic delta precisely cancels the $-2\times 4$ charge of the four Clifford vacua.

Since (for four-point amplitudes) there is a unique solution to the SUSY Ward identities, we can return to the $10$-dimensional perspective and focus on one of the components of the superamplitude, say the four-gluon one. We will be able to obtain any other component by acting with suitable supersymmetry transformations on it. The generic four-gluon amplitude takes the form
\begin{equation}
    A(gggg) = \mathcal F^4(p_i,\epsilon_i) M(s,u)\,,
\end{equation}
where $M(s,u)$ is effectively a simple ``scalar'' amplitude and the prefactor $\mathcal F^4$ contains all the depedence on the polarization vectors. This prefactor is well-known from string theory and SYM amplitudes, and it can be compactly written as \cite{Arkani-Hamed:2022gsa}
\begin{equation}\label{eq:F4def}
    \mathcal F^4(p_i,\epsilon_i) \equiv \big(F_{\mu\nu} F^{\mu\nu}\big)^2 - 4\big(F_{\mu\nu}F^{\nu\alpha} F_{\alpha\beta} F^{\beta\mu}\big)\,,
\end{equation}
where $F_{\mu\nu} \equiv F_{1,\mu\nu} + \cdots + F_{4,\mu\nu}$, with $F_{i,\mu\nu} \equiv p_{i,\mu} \epsilon_{i,\nu} - p_{i,\nu} \epsilon_{i,\mu}$, and one should keep only terms linear in all four polarization vectors $\epsilon_{i,\mu}$. We note that it is invariant under permutations of the external legs. The explicit prefactors for the four-point amplitudes involving the remaining components of the supermultiplet (which are fixed by supersymmetry) can be found, for example, in \cite{Green:1987sp}.

\subsection{$\mathcal N = (1,1)$}
We conclude this  review by commenting on the other maximally supersymmetric case, $\mathcal N=(1,1)$, with two supercharges $Q^A,\widetilde Q_{A'}$ of different chirality. In this case, the massless supermultiplet is composed of the graviton, dilaton, $B$-field, RR forms of odd rank, and two gravitinos of opposite chirality. On this representation, the supersymmetry algebra amounts to two independent copies of the $\mathcal N=(1,0)$ algebra. Unlike the case of $\mathcal N=(2,0)$, since the two supercharges have different chirality, we cannot arrange them into raising and lowering operators while preserving explicit $SO(8)$ covariance. This means that we cannot construct a superspace analogous to the one for $\mathcal N=(2,0)$. Instead, we can repeat the trick used above for $\mathcal N=(1,0)$ of reducing to four dimensions.

In this case, this restricts to a four-dimensional $\mathcal N=8$ SUSY algebra. The ten-dimensional multiplet becomes the familiar graviton multiplet with maximal supersymmetry,
\begin{equation}
    \Phi(p,\eta) = \varphi_- + \eta_I \psi^I + \frac{1}{2!}\eta_I\eta_J\varphi^{IJ} + \cdots + \eta_1\ldots\eta_8\, \varphi_+\,,
\end{equation}
and the four-point amplitude has again a solution in terms of a fermionic delta function~\cite{Elvang:2009wd,Elvang:2010xn},
\begin{equation}
    \mathcal A_4(p_i,\eta_i) = \delta^{10}({\textstyle\sum_i} p_i)\delta^{16}(\bar Q) M(s,u)\,, \qquad
    \delta^{16}(\bar Q) \equiv \frac{1}{2^8} \prod_{I=1}^8 \bar Q_{I\dot \beta} \bar Q_{I}^{\dot \beta}\,.
\end{equation}
We see that this solution is again unique because the fermionic delta has $R$-charge $+16$ and each Clifford vacuum contributes $-4$. In terms of ten-dimensional amplitudes, the various prefactors are simply the ``square'' of the $\mathcal N = (1,0)$ prefactors, as is famously the case in string theory \cite{Kawai:1985xq}. In particular, four-graviton scattering involves a factor $\mathcal R^4 \equiv \left(\mathcal F^4\right)^2$ with the same $\mathcal F^4$ defined in \eqref{eq:F4def}. From the crossing properties of $\mathcal R^4$, we conclude that $M(s,u)$ is again fully crossing symmetric, as in the $\mathcal N=(2,0)$ case.

\section{Setup for supergraviton scattering}\label{sec:closedStringsSetup}
We have just reviewed how, in $D=10$ with maximal supersymmetry (i.e.\ both with $\mathcal N =(2,0)$ and $\mathcal N = (1,1)$), the four-point scattering amplitudes of the graviton supermultiplet are completely determined in terms of a single auxiliary scalar amplitude $M(s,u)$ that is fully crossing symmetric. We will now discuss the general properties of this basic amplitude. We defer the analogous discussion for $\mathcal N =(1,0)$ SYM amplitudes to section \ref{sec:openStrings}.

\subsection{Analyticity, unitarity, Regge behavior}
We focus on  tree-level extensions of supergravity.
As a consequence,  $M(s,u)$ is analytic anywhere on the complex $s$ plane (at fixed $u$) away from the real axis; its only singularities are for real $s$, from the tree-level exchanges of physical states in the theory.
A given exchange in the basic amplitude $M(s,u)$ has different space-time interpretations depending on the component of the superamplitude $\mathcal A_4$ that we are looking at.
That is, different amplitudes in $\mathcal A_4$ exchange different components of a given supermultiplet, but supersymmetry forces their on-shell couplings to be the same.

For a $2 \to 2$ tree-level amplitude of identical particles, unitarity is the statement that all on-shell couplings squared are non-negative. This is expressed in terms of a partial wave expansion
\begin{equation}\label{eq:PWE}
    {\rm Im} \,  M(s,u)  = s^{\frac{4-D}{2}} \sum_{J=\text{even}} n^{(D)}_J \rho_J(s)\, \legP_J\left( {1+\frac{2u}{s}} \right) \,,
\end{equation}
as positivity of the spectral density, 
\begin{equation}\label{eq:positivity}
    \rho_J(s) \geq 0\,, \quad \text{in the physical region} \quad s> 0,\, u<0\,.
\end{equation}
Here and throughout, we use conventions from \cite{Correia:2020xtr} where the normalization constant is
\begin{equation}
    n_J^{(D)}\equiv \frac{2^{D} \pi^{\frac{D-2}{2}}}{\Gamma(\tfrac{D-2}{2})} (J+1)_{D-4} (2J+D-3)\,,
\end{equation}
and the Gegenbauer polynomials are defined by
\begin{equation}\label{eq:Gegenbauer}
    \legP_J(x) \equiv {}_2F_1\left(-J,J+D-3,\frac{D-2}{2},\frac{1-x}{2}\right)\,.
\end{equation}
While we keep expressions general, we will set the space-time dimension to $D=10$ in explicit calculations.

It is important to note that unitarity is a priori a condition on the physical amplitudes contained in $\mathcal A_4$ rather than $M(s,u)$ itself. In the case of $\mathcal N = (2,0)$, the partial wave expansion in \eqref{eq:PWE} is equivalent to the $s$-channel partial wave expansion of the axi-dilaton amplitude \eqref{eq:axi-dilaton} upon a (positive) rescaling of the spectral density by $s^4$. So \eqref{eq:positivity} is equivalent to the positivity of $A(\varphi\varphi\bar\varphi \bar\varphi)$. This is clearly a necessary condition, but it is in fact expected to be also sufficient to guarantee positivity for all other amplitudes in $\mathcal A_4$ \cite{Guerrieri:2021ivu,Boels:2012ie,Bern:1998ug}.\footnote{As a non-trivial check, we note that the crossed axi-dilaton amplitude $A(\varphi\bar\varphi\bar\varphi \varphi) = t^4 M(s,u)$ also has a positive partial wave expansion as long as \eqref{eq:positivity} is satisfied, because $(-s-u)^4 \legP_J\left( {1+\frac{2u}{s}} \right)$ admits a positive expansion in Gegnbauer polynomials. This example allows us to highlight the fact that the spins $J$ appearing in \eqref{eq:PWE} do not translate directly into the spins of the particles exchanged in the physical amplitudes. A scalar pole $\sim \lambda^2 \frac{1}{m^2-s}$ in $M(s,u)$ describes the exchange of a scalar in $A(\varphi\varphi\bar\varphi \bar\varphi)$, but a collection of spins (in the same massive supermultiplet) ranging from 0 through 4 in $A(\varphi\bar\varphi\bar\varphi \varphi)$.} In the case of $\mathcal N =(1,1)$, the same expansion \eqref{eq:PWE} is reached by factoring out the $\mathcal R^4$ factor from the four-graviton amplitude \cite{Arkani-Hamed:2022gsa}.

We have already discussed in the introduction
our assumption (v) of $J<2$ Regge behavior. This is  a statement about physical amplitudes,
e.g.~for the axi-dilaton amplitude $A(\varphi\varphi\bar\varphi \bar\varphi)$.
In terms of the auxiliary amplitude, this translates
into $J < -2$,
\begin{equation}\label{eq:Regge}
\lim_{|s| \to \infty} s^2 M(s, u) \to 0 \quad {\rm for \;fixed}\; {u<0}\,.
    %\lim_{|s| \to \infty} s^2 M(s, u) = 0 \, ,  \quad u < 0 \,.
\end{equation}
Supersymmetry has first related all the amplitudes in the supermultiplet to a single fully-crossing-symmetric scalar amplitude $M(s,u)$, which we now see that it has a Regge behavior improved by four units. This will allow us to write down twice-anti-subtracted dispersion relations, making bootstrap bounds very constraining. In particular, we will be able to write down a sum rule for the Newton constant $G$, equation
(\ref{eq:k=-2}) below. If we only required $O(s^2)$ Regge 
behavior and not better (for the physical amplitude), the Newton constant would not appear in any bootstrap equation.

\subsection{Low energies: Corrections to supergravity}\label{sec:lowEsugra}

At low energies, the scattering of the $\mathcal N = (2,0)$ graviton supermultiplet is dictated by type IIB supergravity. In turn, the case with $\mathcal N = (1,1)$ is given by type IIA. In either case, the supergravity amplitude reads
\begin{equation}\label{eq:Msugra}
    M_{\text{sugra}}(s,u) = \frac{8\pi G}{stu}\,,
\end{equation}
where $G$ is the ten-dimensional Newton constant.
In the type IIB axi-dilaton scattering amplitude \eqref{eq:axi-dilaton}, for example, this simply corresponds to a graviton pole in the $t$ and $u$ channels. More generally, it captures the exchange of the whole massless multiplet in the superamplitude $\mathcal A_4$. Note that \eqref{eq:Msugra} fails to satisfy the Regge behavior \eqref{eq:Regge}, and so it is ruled out by our assumptions. Excited states must kick in at higher energies to UV complete it. The full theory will then be parametrized by a collection $\{m_i^2,J_i,\lambda_{ijk}\}$ of masses, spins and on-shell couplings for said excited states.\footnote{Given the improved Regge behavior \eqref{eq:Regge} of supergraviton amplitudes, one can easily show that three-point couplings are enough to specify the theory using un-subtracted dispersion relations for the four-point amplitude. 
} In general, the spectrum of excited states could be discrete, or continuous, or have accumulation points, etc. We will assume that there is \textit{a gap} $M$ between the (massless) graviton and the first excited state.

The mass gap $M$ defines a cutoff scale below which 
we may use an effective field theory description where we have integrated out all of the massive states. The resulting low-energy action is given by supergravity plus higher-derivative corrections parametrized by (unknown) Wilson coefficients $g_i$. 
At the level of the amplitude, the existence of a gap implies that the only singularity in $M(s,u)$ around the origin is the graviton pole, and we can therefore use a (crossing-symmetric) polynomial expansion
\begin{equation}\label{eq:Mlow}
    M_{\rm low}(s,u) = \frac{8\pi G}{stu} + g_0 + g_2(s^2 + t^2 + u^2) + g_3\, stu + \ldots\,,
\end{equation}
with radius of convergence dictated by $M^2$. Each of these additional terms can be mapped to a higher-derivative correction to the supergravity action. For example, the $g_0$ term corresponds to an $\sim R^4$ correction \cite{Wang:2015jna}.\footnote{Lower powers of the Riemann tensor are forbidden by supersymmetry. Indeed, the only other term of lower mass dimension that is compatible with the symmetries of  $M(s,u)$ is $1/s + 1/t + 1/u$, but this would correspond to  massless states of higher spin \cite{Guerrieri:2021ivu}.}
The infinite set of Wilson coefficients $\{g_i\}$ provides an alternative parametrization for the putative UV 
tree-level completions of supergravity ---all $g_i$ could be computed explicitly if the on-shell data $\{m_i^2,J_i,\lambda_{ijk}\}$ of the underlying theory were known.

In this work, we will proceed to carve out allowed regions in the space of tree-level completions of supergravity by demanding compatibility with the basic requirements that we have reviewed. We will do so both in the space of Wilson coefficients $\{g_i\}$ and of on-shell data $\{m_i^2,J_i,\lambda_{ijk}\}$ (and combinations thereof). In order to access the latter, we will need to refine our low-energy description. Namely, following \cite{Albert:2022oes,Albert:2023seb}, we will raise the cutoff $M^2$ and integrate back in the states with $m_i^2<M^2$. Notice that this involves making further assumptions about the spectrum of low-lying states beyond the existence of a gap. We will come back to these assumptions in section \ref{sec:Resonance}, but for now let us illustrate this in the simplest scenario where the lowest-lying state is a single exchange of mass $m_\phi$ and ``spin'' $J_\phi$.
In this case, we would use a refined effective amplitude containing an explicit pole,
\begin{equation}\label{eq:MlowPole}
    M_\text{low}^{\phi-\text{pole}}(s,u) = \frac{8\pi G}{stu} + \lambda_{\phi}^2\left(\frac{\legP_{J_\phi}\left(1+\frac{2u}{m_\phi^2}\right)}{m_\phi^2 - s} + \frac{\legP_{J_\phi}\left(1+\frac{2s}{m_\phi^2}\right)}{m_\phi^2 - t} + \frac{\legP_{J_\phi}\left(1+\frac{2t}{m_\phi^2}\right)}{m_\phi^2 - u}\right) + \text{analytic} \,,
\end{equation}
which is a valid description up to the mass of the heavier exchanged states. This gives us explicit control over the on-shell coupling $\lambda_\phi^2$ as well as the mass and ``spin'' of the excited state. As a reminder, the ``spin'' $J_\phi$ in \eqref{eq:MlowPole} describes full massive supermultiplets. $J_\phi=0$, for instance, corresponds to the smallest such multiplet, with $SO(9)$ little group representations
\begin{equation}\label{eq:SO(9)rep}
    (\mathbf{44} + \mathbf{84} + \mathbf{128})\otimes (\mathbf{44} + \mathbf{84} + \mathbf{128})\,.
\end{equation}

\subsection{The example: Type II string theory}\label{sec:stringthy}
The only known fully-consistent 
 extensions of the  non-chiral and chiral
maxinal supergravities are respectively the Type IIA and IIB superstring theories. At tree level, the basic amplitude  $M(s,u)$ for {\it both}  IIA and IIB is given by the Virasoro-Shapiro amplitude \cite{Virasoro:1969me,Shapiro:1970gy,Green:1987sp}
\begin{equation}\label{eq:MST}
    M_{\text{ST}}(s,u) = -\frac{\Gamma\left(\hbox{$-\frac{\alpha' s}{4}$}\right) \Gamma\left(\hbox{$-\frac{\alpha' t}{4}$}\right) \Gamma\left(\hbox{$-\frac{\alpha' u}{4}$}\right)}{\Gamma\left(\hbox{$1+\frac{\alpha' s}{4}$}\right) \Gamma\left(\hbox{$1+\frac{\alpha' t}{4}$}\right) \Gamma\left(\hbox{$1+\frac{\alpha' u}{4}$}\right)}\,.
\end{equation}
This amplitude, of course, satisfies all of our assumptions. It is meromorphic with only physical real singularities,
fully crossing symmetric, and it grows as $M_{\text{ST}}(s,u)\sim s^{-2+2\alpha' u}$ in the Regge limit, making it compatible with \eqref{eq:Regge}. Unitarity in the sense of \eqref{eq:positivity} is difficult to prove for all residues (although see \cite{Arkani-Hamed:2022gsa} for recent progress in this direction), but it can be shown indirectly using the no-ghost theorem of worldsheet string theory \cite{Goddard:1972iy,Aoki:1990yn}.

By expanding \eqref{eq:MST} at low energies $s,u \sim 0$ and comparing to \eqref{eq:Mlow} we can extract the $\alpha'$ corrections of string theory to the supergravity amplitude,
\begin{equation}
    \text{ST:}\qquad 8\pi G = \frac{64}{\alpha'^3}\,, \quad
    g_0 = 2\zeta(3)\approx 2.40411...\,, \quad
    g_2 = \frac{\alpha'^2}{16} \zeta(5) \approx \frac{\alpha'^2}{16} 1.03693...\,.
\end{equation}
Since the first massive pole in \eqref{eq:MST} sits at mass $m_\phi^2 = 4/\alpha'$, this is a valid expansion below a cutoff $M^2$ as long as $ M^2 \leq 4/\alpha'$. For energies beyond this, we need to refine our EFT to include the pole explicitly. The first massive string mode is in the supermultiplet \eqref{eq:SO(9)rep}, so we can use the expansion \eqref{eq:MlowPole} with $J_\phi=0$, which is then valid up to a higher cutoff $4/\alpha' \leq M^2 \leq 8/\alpha'$. By computing the residue we can extract the on-shell data
\begin{equation}
    \text{ST:}\qquad m_\phi^2 = 4/\alpha'\,, \quad J_\phi=0\,,\quad \lambda_\phi^2 = 4/\alpha'\,.
\end{equation}
We will later compare these values compare to our bounds.

\subsection{Positivity bounds from dispersion relations}\label{sec:posbounds}
Having discussed the general form of supergraviton amplitudes, their low-energy limit, and the known UV completion from string theory, we now turn to set up our bootstrap problem. The main question being: is string theory the only possible (tree-level) UV completion of supergravity? To address this question, we follow the blueprint of \cite{Caron-Huot:2020cmc} for deriving optimal bounds for low-energy EFT couplings, subsequently upgraded in \cite{Caron-Huot:2021rmr} to handle massless poles. Since this has become a standard procedure, we will be brief in our review, referring the reader to the original references for an introduction, and deferring the details of the numerical implementation to appendix~\ref{app:numerics}.

We start by writing down dispersion relations (with $u<0$),
\begin{equation}
    \frac{1}{2\pi i} \oint_\infty \frac{ds}{s} \frac{M(s,u)}{[s(s+u)]^{k/2}} = 0\,, \qquad k = -2,0,2,4,\ldots\,,
\end{equation}
which vanish by the Regge behavior \eqref{eq:Regge}. By analyticity, we are then allowed to shrink the contour towards the real axis, obtaining a nontrivial link between the IR and UV poles, separated by the cutoff $M^2$,
\begin{equation}\label{eq:IR-UV}
    \frac{1}{2\pi i} \oint_{s\sim M^2} \frac{ds}{s} \frac{M(s,u)}{[s(s+u)]^{k/2}} = \avg{\frac{2m^2+u}{m^2+u} \frac{\legP_J(1+\tfrac{2u}{m^2})}{[m^2(m^2+u)]^{k/2}}}\, .
\end{equation}
Here, following \cite{Caron-Huot:2020cmc}, we have defined the high-energy average by
\begin{equation}\label{eq:heavyavg}
    \avg{(\cdots)} \equiv \frac{1}{\pi} \sum_{J \text{ even}}  n^{(D)}_J
 \int_{M^2}^\infty \frac{dm^2}{m^2} m^{4-D}  \rho_J(m^2) \;(\cdots)\, .
\end{equation}
Since the integral on the left hand side of \eqref{eq:IR-UV} involves only low energies, we can now plug in it the EFT expansion \eqref{eq:Mlow} to express the various EFT data in terms of high-energy averages.

Explicitly, the first few dispersion relations read
\begin{subequations}
\begin{alignat}{2}\label{eq:k=-2}
    &k=-2 \hspace{95pt} -\frac{8\pi G}{u} \,&&= \avg{m^2 (2m^2+u) \legP_J(1+\tfrac{2u}{m^2})}\,, \\\label{eq:k=0}
    &k=0 \qquad g_0 + 2g_2u^2 + 4 g_4 u^4 + \cdots \,&&= \avg{\frac{2m^2+u}{m^2+u} \legP_J(1+\tfrac{2u}{m^2})}\,, \\\label{eq:k=2}
    &k=2 \qquad \hspace{5pt} 2g_2-g_3u+8g_4 u^2+ \cdots \,&&= \avg{\frac{2m^2+u}{m^2+u} \frac{\legP_J(1+\tfrac{2u}{m^2})}{m^2(m^2+u)}}\,,\\
    & &&\cdots\,. \nonumber
\end{alignat}
\end{subequations}
For the dispersion relations with $k\geq 0$, we can follow the usual prescription \cite{Caron-Huot:2020cmc} of expanding around the forward limit $u\sim 0$ and matching coefficients. In this way we can extract an infinite set of \textit{sum rules} expressing each low-energy coupling $g_i$ as a high-energy average. The ones of most interest to us will be
\begin{equation}\label{eq:g0g2avg}
    g_0 = \, \avg{2}\,, \qquad
    g_2 = \, \avg{\frac{1}{m^4}}\,.
\end{equation}

The case of $k=-2$ is special. Due to the massless pole of the graviton supermultiplet, this dispersion relation cannot be expanded in the forward limit.\footnote{The fact that the massless pole is disentangled from the contact terms is a consequence of the improved Regge behavior granted by supersymmetry. In the non-supersymmetric case, one needs to ``improve'' the dispersion relations by hand, making use of other forward-limit sum rules \cite{Caron-Huot:2021rmr}.} A way around this issue was introduced in \cite{Caron-Huot:2021rmr}. The idea is to \textit{smear} the pole by integrating the dispersion relation in $u=-p^2$ over its region of validity $p\in (0,M)$. More concretely, we can obtain a set of sum rules for the gravitational coupling $G$ by integrating \eqref{eq:k=-2} against a set of wavepackets $f_k(p)$,
\begin{equation}\label{eq:Gsmeared}
    8\pi G \int_0^{M} \hspace{-5pt}dp \, \frac{f_k(p)}{p^2} = \avg{\int_0^{M}\hspace{-5pt} dp \, f_k(p)\, m^2 (2m^2 - p^2) \legP_J(1-\tfrac{2p^2}{m^2})}\,.
\end{equation}
There is some leeway in the choice of these wavepackets. For spacetime dimension $D=10$, a particularly nice set is given by $f_k(p) = \{p^{3/2},p^{5/2},p^{7/2},... \}$ \cite{Caron-Huot:2021rmr}. This set offers a technical advantage in that it simplifies the asymptotics of the right hand side of \eqref{eq:Gsmeared}.

As usual, crossing symmetry is encoded in terms of \textit{null constraints}, first introduced in \cite{Caron-Huot:2020cmc,Tolley:2020gtv}. These come from sum rules for low-energy quantities that are set to zero by crossing symmetry. The lowest such constraint comes from the $O(u^1)$ term in \eqref{eq:k=0}, the next one comes from subtracting the two sum rules for $g_2$ from \eqref{eq:k=0} and \eqref{eq:k=2}, etc.
All null constraints can be systematically computed from the double contour integral \cite{Albert:2023jtd}
\begin{equation}\label{eq:NCint}
    \oint_0 \frac{du}{2\pi i}\oint_\infty \frac{ds}{2\pi i}\frac{1}{s u}\left(\frac{F(s,u)}{s^{n-\ell}u^\ell} - \frac{F(u,s)}{u^{n-\ell}s^\ell}\right)=0\,, \quad \begin{cases}\text{for }n-\ell\geq 0\,,\\ \text{and }\ell\geq 0\,,\end{cases}
\end{equation}
for $F(s,u)=M(s,u)$.
The antisymmetry of the kernel kills all low-energy contributions and, upon deforming the contour, we directly reach nontrivial constraints for the high-energy data in the form $\avg{\mathcal X_{n,\ell}}=0$, where
\begin{equation}
    \mathcal X_{n,\ell} = \mathop{\mathrm{Res}}_{u = 0}
    \left[\frac{1}{u}\left(\left(\frac{1}{m^{2(n-\ell+1)}u^\ell} - \frac{1}{u^{n-\ell}m^{2\ell+2}}\right)
    -(m^2\rightarrow -m^2-u)
    \right)m^2\legP_J(1+\tfrac{2u}{m^2})
    \right]\,.
\end{equation}
The set of independent null constraints is given by $n=1,2,...$, $\ell=0,1,...,\left[\frac{n-1}{3}\right]$.

Equipped with sum rules and null constraints, we can arrange the problem of deriving bounds for ratios of on-shell and EFT couplings into a semidefinite program \cite{Caron-Huot:2020cmc,Caron-Huot:2021rmr}, which can then be solved e.g.\ with \texttt{SDPB} \cite{Simmons-Duffin:2015qma}. As an example, to derive a bound on $\frac{g_0 M^6}{8\pi G}$, we define the vector
\begin{equation}
    \Vec v_{\text{HE}}\equiv \left(-G_1(m^2,J),\ldots, -G_{k_\text{max}}(m^2,J), -2M^6, \mathcal X_{1,0}(m^2,J), \cdots ,\mathcal X_{n_{\text{max}},\left[\frac{n_{\text{max}}-1}{3}\right]}(m^2,J) \right)\,,
\end{equation}
where $G_k(m^2,J)$ is the argument on the right hand side of \eqref{eq:Gsmeared} after smearing against the wavepacket $f_k(p)=p^{k+1/2}$. We truncate the set of wavepackets at order $k_{\text{max}}$ and the number of null constraints at order $n_\text{max}$. Bounds can only strengthen when increasing either of these cutoffs. Under the high-energy average, this vector satisfies the bootstrap equation
\begin{equation}\label{eq:g0BootstrapEq}
    8\pi G \Vec v_{\text{obj}} + g_0 M^6 \Vec v_{\text{norm}} + \avg{\Vec v_{\text{HE}}(m^2,J)}=0\,,
\end{equation}
where we have defined the vectors
\begin{align}
    \Vec v_{\text{obj}} \,&\equiv \left(2 M^{1/2},\ldots, \tfrac{M^{k_\text{max}-1/2}}{k_\text{max}-1/2}, 0,0, \ldots,0\right)\,,\\
    \Vec v_{\text{norm}} \,&\equiv \left(0,\ldots, 0, 1,0, \ldots, 0\right)\,.\nonumber
\end{align}
The bound is then simply obtained by looking for a vector $\Vec \alpha$ such that $\Vec \alpha \cdot \Vec v_{\text{norm}} = 1$, $\Vec \alpha \cdot \Vec v_{\text{HE}}(m^2,J)\geq 0$ for all even $J, m\geq M$, and which maximizes $\Vec \alpha \cdot \Vec v_{\text{obj}}$. We explain in more detail our truncations in $m$ and $J$ for the positivity condition in appendix \ref{app:numerics}.

Solving this semidefinite problem with \texttt{SDPB} \cite{Simmons-Duffin:2015qma} for $k_\text{max} = 31$ and $n_\text{max} = 31$ yields the optimal bound
\begin{equation}\label{eq:g0Bound}
    \widetilde g_0 \equiv \frac{g_0 M^{6}}{8\pi G} \leq 2.969\,.
\end{equation}
This is a minor improvement on the bound found in \cite{Caron-Huot:2021rmr}, where both sum rules and null constraints were implemented through smearing. Mixing the smeared graviton pole with the forward limit higher subtracted EFT sum rules and null constraints is computationally easier and allows us to quickly reach higher $n_{\text{max}}$.

\section{Results for supergravitons}\label{sec:CSResults}
\subsection{EFT bounds}\label{sec:CSeftbounds}
\begin{figure}[ht]
\centering
\includegraphics[width=0.8\textwidth]{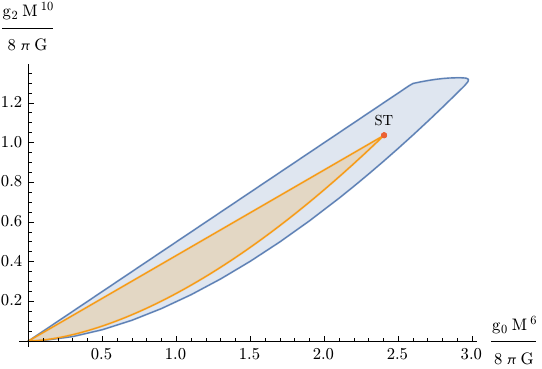}
\caption{Exclusion plot in the space of the first two EFT coefficients, $g_0$ and $g_2$, normalized by $G$ and the cutoff $M$. The (blue) shaded region is the allowed space for amplitudes. It was generated numerically using $k_{\text{max}}=n_{\text{max}}=29$ (i.e. a total of 14 wavepackets and 155 null constraints). The red dot corresponds to the Virasoro-Shapiro amplitude \eqref{eq:MST} with $M^2\alpha'/4=1$. Varying $M^2\alpha'/4$ from 0 to 1 rules in the orange region.}
\label{fig:CSg0g2}
\end{figure}

Having found an upper bound on $\widetilde g_0$ one can ask what the space of allowed amplitudes looks like when considering multiple Wilson coefficients in combination. We can add the next coefficient, $g_2$, into the mix by modifying the bootstrap equation:
\begin{equation}
    8\pi G\Vec v_{\text{obj}} + g_2M^{10} \Vec v_{\text{norm}} + \avg{\Vec v_{\text{HE}}(m^2,J)}=0\,,
\end{equation}
where now
\begin{align}
    \Vec v_{\text{HE}}\,& \equiv \left(-G_1(m^2,J),\ldots, -2M^6, -M^{10}/m^4, \mathcal X_{1,0}(m^2,J), \cdots \right)\,,\nonumber\\
    \Vec v_{\text{obj}} \,&\equiv \left(2 M^{1/2},\ldots, \widetilde g_0,0,0, \ldots\right)\,,\\\
    \Vec v_{\text{norm}} \,&\equiv \left(0,\ldots, 0, 1,0, \ldots\right)\,.\nonumber
\end{align}
For fixed $\widetilde{g}_0$ one now can get an upper and lower bound on $\widetilde{g}_2\equiv \frac{g_2 M^{10}}{8\pi G}$ (normalizing by $\Vec \alpha \cdot \Vec v_{\text{norm}} = 1$ and $-1$ respectively). Scanning over the allowed range of $\widetilde{g}_0$ then generates an exclusion plot, see figure \ref{fig:CSg0g2}.

We find some interesting features: Up to $\widetilde{g}_0\approx 2.6$ the upper bound on $\widetilde{g}_2$ lies exactly on the line $\widetilde{g}_2=\frac{1}{2}\widetilde{g}_0$. This is the bound that we could have derived if we had just disregarded all null constraints. Since $\frac{m^2}{M^2}\geq 1$ we must have $\avg{\frac{M^4}{m^4}}\leq\frac{1}{2}{\avg{2}}$ which comparing to \eqref{eq:g0g2avg} gives $\widetilde{g_2}\leq \frac{1}{2}\widetilde{g}_0$. In order to saturate this bound, the extremal amplitude must have a tower of states at the cutoff and no states at any finite mass above. This would be achieved by the following amplitude:
\begin{equation}\label{eq:Mtower}
    M_{\text{tower}}(s,u) = \frac{8\pi G}{stu} + \frac{\chi}{(M^2-s)(M^2-t)(M^2-u)}\,,
\end{equation}
which yields $\widetilde g_0 = 2\widetilde g_2 = \chi/8\pi G$. This amplitude is in principle not quite allowed because it grows exactly like $M_{\text{tower}}(s,u)\sim s^{-2}$ in the Regge limit, and we imposed a Regge behavior \textit{strictly better} than this (c.f.\ \eqref{eq:Regge}). But since its growth is only borderline disallowed, we expect that it can be cured by adding states at a parametrically high mass, in analogy to the spin-one amplitudes discussed in \cite{Albert:2022oes}.\footnote{These modifications leave the low-energy coefficients unchanged, but the allowed range for $\chi$ depends on the precise completion. It would be interesting to find a completion that saturates the upper bound $\chi/8\pi G\lesssim 2.6$.}
Above $\widetilde{g}_0\approx 2.6$ the upper bound on $\widetilde{g}_2$ flattens and connects to an extremal point that maximizes both $\widetilde g_0$ and $\widetilde g_2$. An interpretation of this, as well as of the lower curved bound will only become clear in subsequent sections.

Let us look at where the string theory amplitudes \eqref{eq:MST}  fall compared to our bounds. Note that rather than a point in theory space, they comprise a continuous family parametrized by $\alpha'$. These amplitudes are compatible with our assumptions as long as the first pole, $s=4/\alpha'$, is above the cutoff $M^2$; so $M^2\alpha'/4\in (0,1]$. In $\widetilde{g}_0$-$\widetilde{g}_2$ space, varying $M^2\alpha'/4$ in this range traces the curve
\begin{equation}
    \widetilde g_0 = 2\zeta(3) \left(\frac{M^2 \alpha'}{4}\right)^3\,, \qquad
    \widetilde g_2 = \zeta(5) \left(\frac{M^2\alpha'}{4}\right)^5\,.
\end{equation}
In fact, by considering (positive) linear combinations of Virasoro-Shapiro amplitudes with varying $\alpha'$ (in the allowed range), we can cover the entire convex hull of this curve ---the orange region in figure \ref{fig:CSg0g2}.
The red dot in figure \ref{fig:CSg0g2} marks the extremal point $M^2\alpha'/4=1$, where the first resonance has mass equal to the cutoff. The other end of the curve, $M^2\alpha'/4\to 0$, tends to pure supergravity \eqref{eq:Msugra} where all Wilson coefficients are set to zero (compared to~$8\pi G$).

Pure supergravity is inconsistent with the strict Regge behavior assumption \eqref{eq:Regge}, albeit only marginally so. Nevertheless, it is always a limiting point of our exclusion plots, for the following somewhat trivial reason. Given any consistent amplitude (in particular, with Regge behavior strictly better than minus two), let us denote by $m_{\rm phys}$ the mass of the first massive exchanged state. While the natural value of the EFT cut-off is $M = m_{\rm phys}$, we can artificially choose $M \ll m_{\rm phys}$; this pushes to zero all dimensionless EFT coefficients, formally giving the pure sugra answer in the limit $M/m_{\rm phys} \to 0$. The lower end of the orange curve in figure \ref{fig:CSg0g2} realizes this from the string theory solution, but the same point can be reached from any other consistent amplitude. We are interested in tree-level S-matrices where the additional states come in at a finite mass $m_{\rm phys} \equiv M$, i.e.\ the bounds away from the origin in figure \ref{fig:CSg0g2}.

\subsection{Adding a massive resonance}\label{sec:Resonance}

As we just saw, there are amplitudes with infinitely many states of the same mass and arbitrary high spin that saturate some of the bounds on the EFT coefficients. 
Such amplitudes cannot be physical, as they violate locality, and we would like to exclude them. 
We do so by making the following restriction on the spectrum: 
the first massive level should comprise a {\it finite} set of spins.\footnote{At the level of a single $2\to 2$ amplitude, we are unable to distinguish degenerate states with the same mass {\it and} the same spin, which would contribute in exactly the same way to the partial wave expansion.}

We start by allowing only for a single state of ``effective spin'' $J_\phi$ (as viewed from the basic amplitude $M(s,u)$), and will upgrade to a (small) finite number of them momentarily. This is done by considering the refined low-energy amplitude \eqref{eq:MlowPole} with an explicit pole at mass $m_\phi$. This is a valid description up to a higher cutoff $M > m_\phi$, after which we are again agnostic about the spectrum.
Note that with this assumption, we will be restricting to (tree-level) UV completions of supergravity where the first excited state is \textit{isolated}. This is a very mild and eminently reasonable assumption, but it excludes more exotic scenarios with an accumulation point or a cut beginning at $m_\phi$.

In practice, the easiest way to implement this addition to the problem, as was first done in \cite{Albert:2022oes}, is by modifying the high energy side of the sum rules.
The spectral function in \eqref{eq:heavyavg} can be taken to explicitly include the first massive resonance by the following shift:
\begin{equation}\label{eq:rhoShift}
    n_{J}^{(D)}\rho_{J}(m^2)\rightarrow \lambda_{\phi}^2\,\delta_{J,J_\phi}\,\pi\delta(m^2-m_{\phi}^2)m_\phi^{D-2} + n_{J}^{(D)}\widetilde{\rho}_{J}(m^2),
\end{equation}
where $\widetilde{\rho}_{J}(m^2)$ now only has support on $m^2\geq M^2\geq m_{\phi}^2$. Here $\lambda_\phi$ corresponds to the on-shell three-point coupling between two states in the graviton supermultiplet and the massive ``spin $J_\phi$'' state being exchanged. Plugging the above into the definition of the heavy average \eqref{eq:heavyavg} shifts all high-energy averages by
\begin{equation}
    \avg{F(m^2,J)}\longrightarrow \lambda_\phi^2 F(m_\phi^2,J_\phi) + \avg{F(m^2,J)}\,.
\end{equation}
We can now repeat the algorithm from the previous section and derive new bounds for the Wilson coefficients, to which we will return in section \ref{sec:onShell+EFT}.

However, with this setup we gain access to an arguably more interesting observable; the on-shell coupling $\lambda_\phi^2$. To bound it, we consider yet a new bootstrap equation,
\begin{equation}
    8\pi G \Vec v_{\text{obj}} + \lambda_{\phi}^2 \Vec v_{\text{norm}} + \avg{\Vec v_{\text{HE}}(m^2,J)}=0\,,
\end{equation}
where\footnote{Note that we should still smear the graviton pole only up to the first resonance, so the integrals in \eqref{eq:Gsmeared} range in ${p\in (0,m_\phi)}$ in the current conventions.}
\begin{align}\label{eq:CSlowJvec}
    \Vec v_{\text{HE}}\,& \equiv \left(-G_1(m^2,J),\ldots, \mathcal X_{1,0}(m^2,J), \cdots \right)\,,\nonumber\\
    \Vec v_{\text{obj}} \,&\equiv \left(2 m_\phi^{1/2},\ldots,0, \ldots\right)\,,\\\
    \Vec v_{\text{norm}} \,&\equiv \Vec v_{\text{HE}}(m_{\phi}^2,J_\phi)\,.\nonumber
\end{align}
We take $m_\phi$ to set the scale of our problem and view the ratio $M^2/m_\phi^2\in [1,\infty)$ as a parameter that we can dial to specify the gap after the resonance $\phi$. For this reason, from this point onwards, we normalize couplings by $m_\phi$ rather than the cutoff $M$. With this bootstrap equation, we can immediately proceed to place bounds on the ratio of on-shell couplings $\widetilde{\lambda}_{\phi}^2\equiv \frac{m_{\phi}^4}{8\pi G}\lambda_{\phi}^2$.

\begin{figure}[!ht]
\centering
\includegraphics[width=0.8\textwidth]{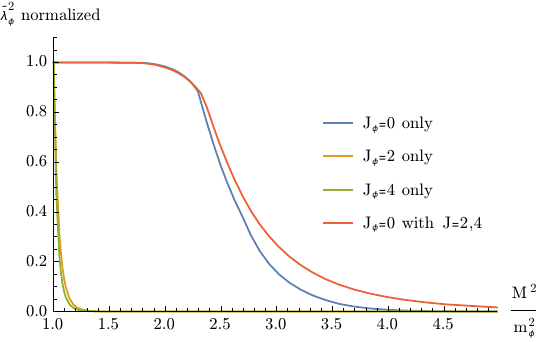}
\caption{Upper bound on the coupling $\widetilde \lambda_\phi^2$ of a single first massive exchanged state of spin 0,2, or 4 (blue, orange, green) as well as the upper bound on the coupling of a spin 0 exchange allowing for additional spin 2 and 4 exchanges. They are all plotted as a function of the cutoff $M^2/m_\phi^2$, describing the mass after which new states may kick in. To aid the comparison, the couplings have been normalized by their respective upper bound at $M=m_{\phi}$. We see that only the $J_\phi=0$ state survives when a gap is imposed and that adding more spins to it does not change the bound by much. This plot was generated with $k_{\text{max}}=n_{\text{max}}=15$ for $J_\phi=2,4$ and $k_{\text{max}}=n_{\text{max}}=13$ for both $J_\phi=0$ bounds.
}
\label{fig:CSdiffSpins}
\end{figure}

In figure \ref{fig:CSdiffSpins} we compare the \textit{upper bound} on this coupling as a function of $M^2/m_\phi^2$ for the cases where the isolated exchange has effective spin $J_\phi=0,2$ and $4$ (blue, orange, and green lines respectively).\footnote{We would like to stress at this point that increasing the cutoff leads to some numerical subtleties. These are discussed in appendix \ref{app:numerics}.} We are normalizing each bound by its maximal value (which happens at $M^2/m_\phi^2=1$) to make the comparison cleaner.
One can immediately see that the bounds on higher spin resonances fall off very quickly with $M^2/m_\phi^2$ (and a sharper decline is expected as one increases $n_{\text{max}}$ and $k_{\text{max}}$). The bound for $J_\phi=0$, in contrast, remains constant for a wide range of the gap $M^2/m_\phi^2$ before decaying. This strongly suggests that, if the first excited state is \textit{isolated}, this state must have $J_\phi=0$, i.e.\ it must be in the smallest massive supermultiplet \eqref{eq:SO(9)rep}.

What if instead of a single state we allow for multiple? We can implement this by taking $\widetilde{\rho}_{J}(m^2)$ to have additional isolated support on the desired spins at $m=m_{\phi}$ (in other words, requiring $\Vec{\alpha}\cdot \Vec{v}_{\text{HE}}\geq 0$ below the cutoff at these points too). In figure \ref{fig:CSdiffSpins} the red line shows the upper bound on a spin 0 state this time also allowing for spins 2 and 4 at the same mass. We see that allowing for these additional states does not change the bound significantly; the plateau remains untouched, and only the tail becomes wider (but this part is still expected to become steeper when increasing $n_\text{max}$ and $k_\text{max}$).
If one looks at the coupling of the spin 2 and 4 states in this setup (not pictured), they fall off quickly again, only slightly slower than when these states were forced to be alone. We expect the same pattern to extend to any finite number of higher spin states, so this sharpens our previous conclusion: if the first massive level is \textit{isolated} and it involves a \textit{finite number} of degenerate spins, it will only contain a single $J_\phi=0$ supermultiplet.

This is of course the case for string theory, where the first massive level corresponds to smallest massive supermultiplet \eqref{eq:SO(9)rep}, 
which has $J_{\rm eff} =0$. 
That the first massive state should 
have $J_{\rm eff} = 0$ is in fact
expected on much more general grounds, as a consequence of Regge theory.
The Regge behavior \eqref{eq:Regge} allows us to use the Froissart-Gribov projection to derive an analytic continuation in spin\footnote{Analytic continuations in spin are the bread-and-butter of Regge theory, see e.g.\ the textbook \cite{Gribov:2003nw} for an introduction.}   of the partial wave coefficients which extends all the way down to the supergraviton (which has $J_{\rm eff}=-2$). This forces the graviton to ``reggeize''.
Barring some
accidental fine tuning,
the first massive state 
should lie on the graviton trajectory and have $J_{\rm eff} = 0$. However, 
it is certainly
reassuring to get this fact as an output of the numerics!

\subsection{The scalar exchange}\label{sec:ScalarExch}
\begin{figure}[ht]
\centering
\includegraphics[width=0.8\textwidth]{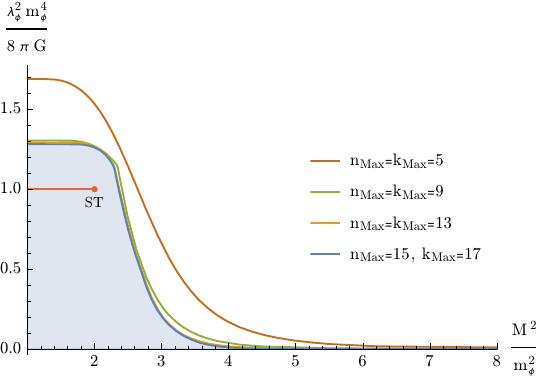}
\caption{Upper bound on the coupling of the first scalar, normalized by its mass and $G$, as a function of the cutoff for various numbers of null constraints and wavepackets. As $n_{\text{max}}$ and $k_{\text{max}}$ are increased, the bound quickly converges. The highest number of null constraints and wavepackets used (blue curve) is $k_{\text{max}}=17$ and $n_{\text{max}}=15$ (i.e. a total of 8 wavepackets and 45 null constraints). 
From unitarity, the lower bound  for all cutoffs is $\widetilde{\lambda}_\phi^2\geq0$. The allowed space of amplitudes is therefore between the upper bound curve and the horizontal axis (area shaded in blue). The Virasoro-Shapiro amplitude \eqref{eq:MST} with $M^2\alpha'=2$ is marked by the red dot; the red line is spanned by choosing an ``unnatural'' cutoff in $1/\alpha' \leq M^2 <2/\alpha'$.}
\label{fig:CSgPM1}
\end{figure}

We have just seen that if the first excited state is isolated, it must be an effective scalar (meaning $J=0$ in the superamplitude \eqref{eq:PWE}). Let us look closer at this exchange.
We begin by looking at the absolute bounds on its squared on-shell coupling $\widetilde{\lambda}_{\phi}^2 \equiv\frac{m_{\phi}^4}{8\pi G}\lambda_{\phi}^2$, which occur at $M^2=m_{\phi}^2$. We find 
\begin{equation}\label{eq:gphirange}
    0\leq \widetilde{\lambda}_{\phi}^2 \leq 1.272\quad\quad (n_{\text{max}}=23,\,k_{\text{max}}=23),\hspace{-4cm}
\end{equation}
the lower bound simply being the one imposed by unitarity. We can now start to push the cutoff higher.
Since the position of the cutoff indicates the earliest point a new state can appear (i.e. no states are allowed to appear with mass $m_{\phi}\leq m < M$), the upper bound must decrease monotonically as the cutoff is increased. The upper bound on $\widetilde{\lambda}_{\phi}^2$ as a function of $M^2/m_\phi^2$ is plotted in figure \ref{fig:CSgPM1}, which is the un-normalized version of the $J_\phi=0$ curve in figure \ref{fig:CSdiffSpins}. The different colors indicate the convergence in the number of wavepackets and null constraints.

Interestingly, as we increase the cutoff $M^2$ from 1 to 2 (in units of the scalar mass) the upper bound stays fixed. Once the cutoff passes 2, the bound quickly drops down to zero. The amplitude that maximizes $\widetilde{\lambda}_{\phi}^2$ must therefore have one or more states at $m^2 \cong 2m_{\phi}^2$. As was discussed in section \ref{sec:stringthy}, this too is the case for string theory. Its first excited state (at mass $m_\phi^2=1/\alpha'$) is a scalar. Next, at mass $2/\alpha'$ it has two states with (effective) spins 0 and 2. The coupling of the first scalar resonance in the Virasoro-Shapiro amplitude, though, gives $\widetilde{\lambda}_{\phi}^2=1$, which is below the upper bound on $\widetilde{\lambda}_{\phi}^2$ (see red dot in figure \ref{fig:CSgPM1}). Note that since we can trivially move the cutoff between $1/\alpha' \leq M^2 \leq 2/\alpha'$, the string theory amplitude actually traces the whole red line in figure \ref{fig:CSgPM1}.

\subsection{Combining on-shell and EFT bounds}\label{sec:onShell+EFT}
We would now like to understand how the restriction to an isolated scalar exchange as the first massive resonance interacts with the bounds on the EFT coefficients. In particular, we are interested in understanding the amplitude that maximizes the scalar coupling $\widetilde{\lambda}_{\phi}^2$. From the discussion above we know that this extremal amplitude does not have states after the first scalar until mass $2m_{\phi}^2$. We therefore set the cutoff to  exactly this point, i.e. $M^2=2m_{\phi}^2$, and proceed to derive bounds on Wilson coefficients.

\begin{figure}[ht]
\centering
\includegraphics[width=0.8\textwidth]{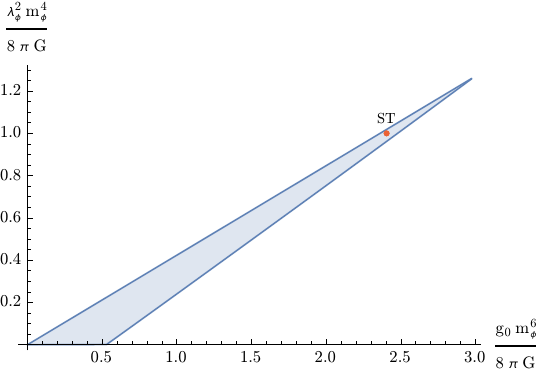}
\caption{Exclusion plot in the space of the first EFT coefficient, $g_0$, and the on-shell coupling of the first massive exchanged scalar, $\lambda_{\phi}$, normalized by $G$ and the mass of the scalar $m_\phi$. The cutoff is set to $M^2=2m_{\phi}^2$. The allowed space of amplitudes is within the (blue) shaded area. The Virasoro-Shapiro amplitude \eqref{eq:MST} is marked by the red dot; it lies within the bounds, relatively close to the boundary. This plot was generated with $k_{\text{max}}=17$ and $n_{\text{max}}=15$ and seems to have converged.}
\label{fig:CSg0gP}
\end{figure}

\begin{figure}[ht]
\centering
\includegraphics[width=0.8\textwidth]{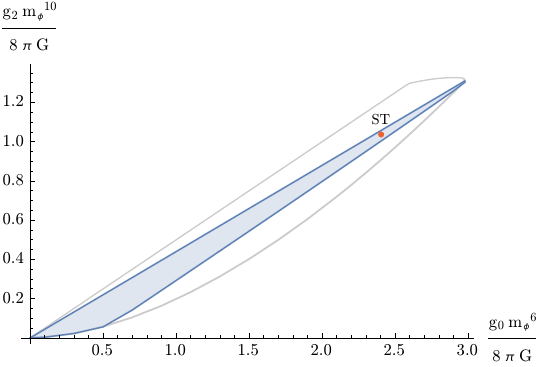}
\caption[{Exclusion plot in the space of the first two EFT coefficients, $g_0$ and $g_2$, with the following spectral assumption: there exists a scalar (and no other states) at mass $m_{\phi}^2$ with on-shell coupling $\widetilde{\lambda}_{\phi}^2\geq 0.001$. The cutoff is set to $M^2=2m_{\phi}^2$. The allowed space of amplitudes is within the (blue) shaded area. The light grey shape shows the original exclusion plot from figure \ref{fig:CSg0g2} without any spectral assumption (i.e. $M^2=m_{\phi}^2$ and $\widetilde{\lambda}_{\phi}^2\geq 0$). The Virasoro-Shapiro amplitude \eqref{eq:MST} is marked by the red dot; it lies within the bounds, relatively close to the boundary. This plot was generated with $k_{\text{max}}=n_{\text{max}}=15$ and seems to have converged.}]
{Exclusion plot in the space of the first two EFT coefficients, $g_0$ and $g_2$, with the following spectral assumption: there exists a scalar (and no other states) at mass $m_{\phi}^2$ with on-shell coupling $\widetilde{\lambda}_{\phi}^2\geq 0.001$\footnotemark. The cutoff is set to $M^2=2m_{\phi}^2$. The allowed space of amplitudes is within the (blue) shaded area. The light grey shape shows the original exclusion plot from figure \ref{fig:CSg0g2} without any spectral assumption (i.e. $M^2=m_{\phi}^2$ and $\widetilde{\lambda}_{\phi}^2\geq 0$). The Virasoro-Shapiro amplitude \eqref{eq:MST} is marked by the red dot; it lies within the bounds, relatively close to the boundary. This plot was generated with $k_{\text{max}}=n_{\text{max}}=15$ and seems to have converged.}
\label{fig:CSg0g2gP}
\end{figure}

\vspace{1mm}
\noindent We look at this system from two angles:
\begin{itemize}
    \item \textbf{Figure \ref{fig:CSg0gP}:} By fixing the scalar coupling, $\widetilde{\lambda}_{\phi}^2$, one can find an upper and lower bound on $\widetilde{g}_0$. Then varying $\widetilde{\lambda}_{\phi}^2$ across its allowed range \eqref{eq:gphirange} generates an exclusion plot coupling the bounds on the first Wilson coefficient to those on the on-shell coupling.

    \item \textbf{Figure \ref{fig:CSg0g2gP}:} Setting a \textit{minimum} value for the scalar coupling (we chose $\widetilde{\lambda}_{\phi}^2\geq 0.001$) enforces the existence of the scalar. With this spectral assumption one can then regenerate the exclusion plot in $\widetilde g_0 = g_0 m_\phi^6/(8 \pi G)$ and $\widetilde g_2 = g_2 m_\phi^{10}/(8 \pi G)$ from section \ref{sec:CSeftbounds} (with the original plot for reference in light grey).
\end{itemize}

\footnotetext{{The value 0.001 was chosen large enough such as to impose the existence of a scalar and avoid just generating a rescaled version of figure \ref{fig:CSg0g2} while small enough to not further restrict the space of the amplitudes. Since string theory has $\widetilde{\lambda}_{\phi}^2=1$ it lies well above the imposed lower bound and hence will not be ruled out by this restriction.}}

What do we learn from these plots? Figure \ref{fig:CSg0gP} is a convex hull spanned by only three points. The amplitude that we found maximizing $\widetilde{\lambda}_{\phi}^2$ is one of them, and we see that it is also the one that maximizes $\widetilde{g}_0$; saturating its most general upper bound from \eqref{eq:g0Bound}.
The second point is an amplitude with $\widetilde{g}_0=\widetilde{\lambda}_{\phi}^2=0$ and the third has vanishing scalar coupling but a non-zero $\widetilde{g}_0 \approx 0.53$. We note that, contrary to what one might expect, this last amplitude is not a simple rescaling of the first one by taking $m_{\phi}^2 \rightarrow 2m_{\phi}^2$.\footnote{Such a rescaling of the first amplitude sits at $\widetilde{\lambda}_{\phi}^2=0$ and $\widetilde{g}_0\cong 2.97/2^3 \approx 0.37$, which is well below the bound.}

These three points appear again in figure \ref{fig:CSg0g2gP}. The $\widetilde{\lambda}_{\phi}^2$-maximizing amplitude sits at the top right extreme of the $\widetilde{g}_0$-$\widetilde{g}_2$ exclusion plot. If we look at the original exclusion plot without spectral assumption (light grey) we can now understand the plateau better. It is generated by the linear combination of the tower of states amplitude \eqref{eq:Mtower} and this new amplitude. The tower of states amplitude in the new exclusion plot (blue) by construction has been ruled out through the spectral assumptions. Continuing to the other two points we find that the amplitude with $\widetilde{\lambda}_{\phi}^2=\widetilde{g}_0=0$ also has $\widetilde{g}_2=0$. The third amplitude ($\widetilde{\lambda}_{\phi}^2=0$ but $\widetilde{g}_0\neq 0$) sits along the lower bound of the original $\widetilde{g}_0$-$\widetilde{g}_2$ plot. The lower bound between these two points follows exactly that of the original (light grey) bound. In the following section we will examine these three extremal points more closely.  

Finally we note that as in the previous plots, the closed string amplitude sits well inside the bulk of the allowed region. In this case the Virasoro-Shapiro amplitude \eqref{eq:MST} is given by a single point because we are using $m_\phi^2 = 4/\alpha'$ to set the scale, and the cutoff has been fixed to $M^2=8/\alpha'$.

\subsection{Analysis of the extremal points}\label{sec:numAnalysis}
\begin{figure}
     \centering
     \begin{subfigure}[b]{0.49\textwidth}
         \centering
         \includegraphics[width=\textwidth]{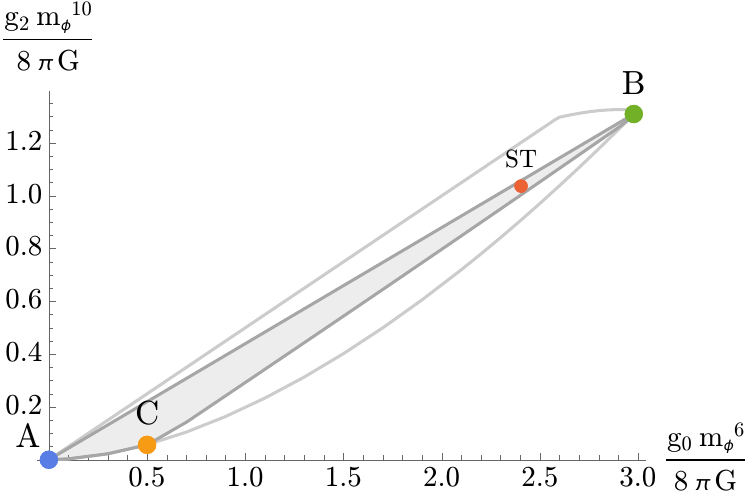}
         \caption{$\widetilde{g}_2$ vs. $\widetilde{g}_0$}
         \label{fig:g0g2color}
     \end{subfigure}
     \hfill
     \begin{subfigure}[b]{0.49\textwidth}
         \centering
         \includegraphics[width=\textwidth]{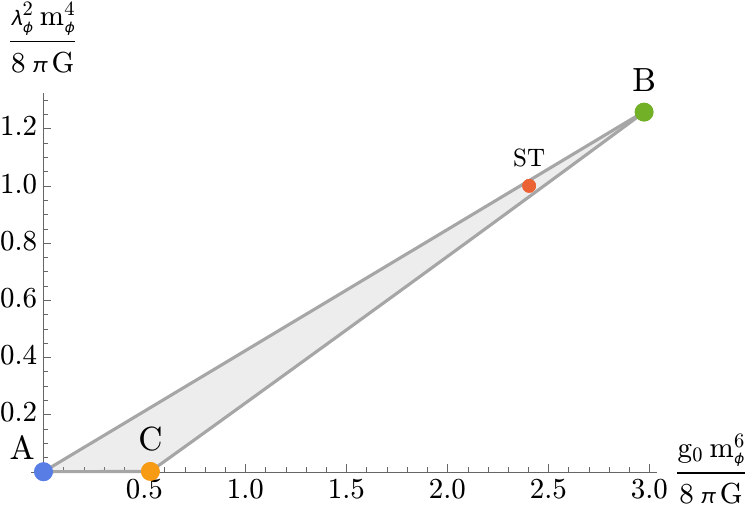}
         \caption{$\widetilde{\lambda}_{\phi}^2$ vs. $\widetilde{g}_0$}
         \label{fig:g0gPcolor}
     \end{subfigure}
     \hfill
     \begin{subfigure}[b]{0.49\textwidth}
         \centering
         \includegraphics[width=\textwidth]{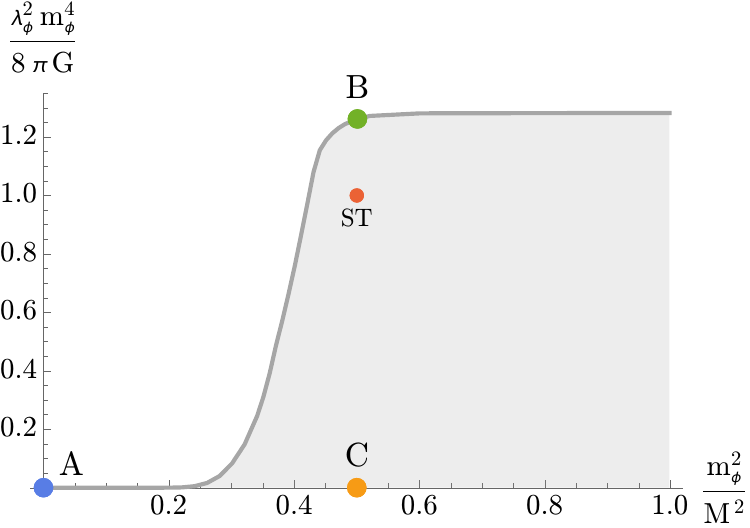}
         \caption{$\widetilde{\lambda}_{\phi}^2$ vs. $m_\phi^2/M^2$}
         \label{fig:g01Mcolor}
     \end{subfigure}
        \caption{Collection of exclusion plots having imposed the first massive resonance (with mass $m_{\phi}$) to have (effective) spin $J_\phi=0$. In \subref{fig:g0g2color} and \subref{fig:g0gPcolor} the cutoff $M^2=2m_{\phi}^2$. In \subref{fig:g01Mcolor} we have inverted the horizontal axis compared to figure \ref{fig:CSgPM1} for visualization purposes. In all plots the red dot marks the Virasoro-Shapiro amplitude \eqref{eq:MST}. The blue (A), green (B), and orange (C) dots denote the position of the three extremal amplitudes.}
        \label{fig:CScolor}
\end{figure}

Our goal in this section is to better understand the three extremal points we found in the exclusion plots above. For reference, figure \ref{fig:CScolor} shows the plots from the previous sections side-by-side with the three points labelled and marked in different colors.

\subsubsection*{Point A}\label{sec:gravitonpole}
Let us start with the easiest one, which we already addressed in section \ref{sec:CSeftbounds}: the amplitude with $\widetilde{\lambda}_{\phi}^2=\widetilde{g}_0=\widetilde{g}_2=0$ (blue points labelled ``A'' in figure \ref{fig:CScolor}).
This point limits to the pure supergravity amplitude \eqref{eq:Msugra}, where all EFT and on-shell couplings are set to zero. We saw in section \ref{sec:CSeftbounds} that this point can be reached by taking the $\alpha'\to 0$ limit of string theory, where all excited states are pushed to infinity.

\subsubsection*{Point B}
\begin{figure}[h]
     \centering
     \begin{subfigure}[b]{0.495\textwidth}
         \centering
         \includegraphics[width=\textwidth]{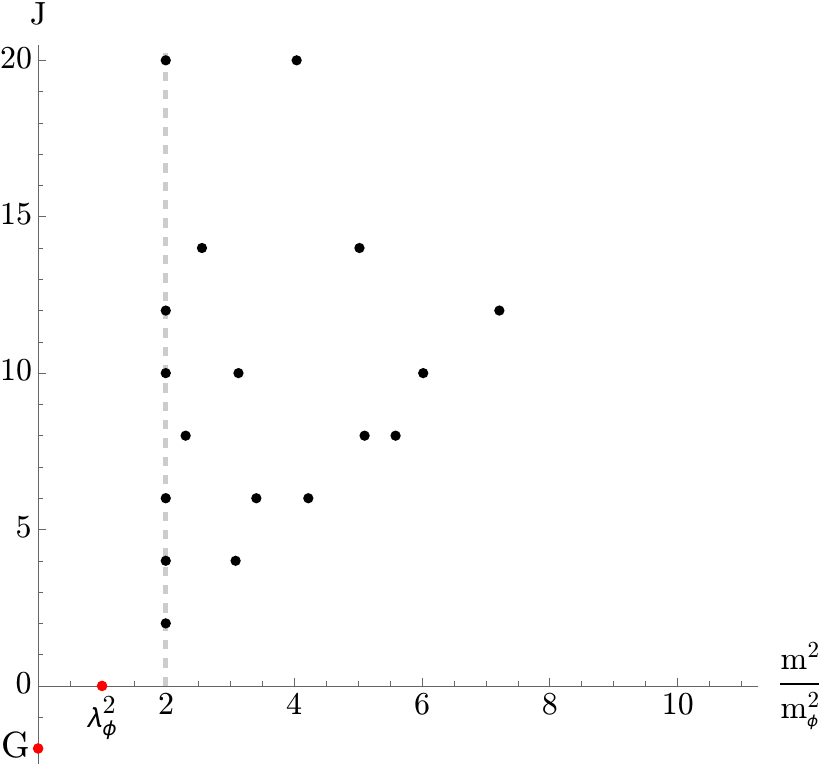}
         \caption{no spectral assumptions}
         \label{fig:singleReggeSnS}
     \end{subfigure}
     \hfill
     \begin{subfigure}[b]{0.495\textwidth}
         \centering
         \includegraphics[width=\textwidth]{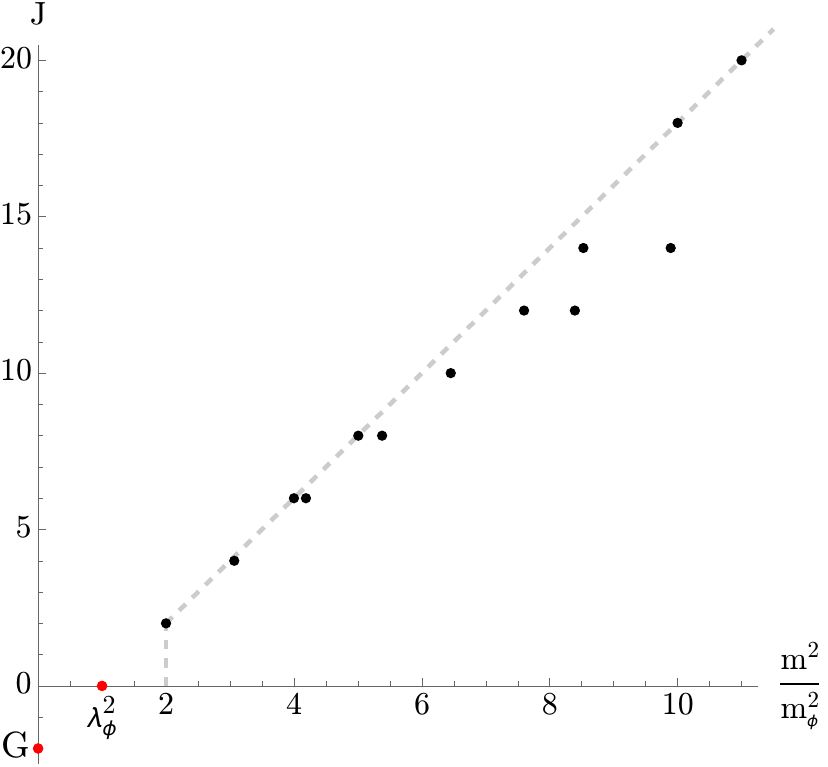}
         \caption{lower triangle assumption}
         \label{fig:singleReggeSwS}
     \end{subfigure}
        \caption{Spectrum outputs from \texttt{SDPB} for the amplitude maximizing the scalar coupling with cutoff at $M^2=2m_{\phi}^2$ (i.e. at Point B). The red dots mark the graviton (which has effective spin $J_{\text{eff}}=-2$) and the enforced massive scalar. The grey dashed line indicates the (possibly spin-dependent) cutoff. (For spins $J>60$ the cutoff $M^2=2m_{\phi}^2$ is used.) Positivity is enforced for all $m^2$ to the right of it. The spectrum \subref{fig:singleReggeSnS} was generated without any additional assumptions on the spectrum.
        States lying above the linear Regge trajectory are spurious since
        excluding them does not change the bound on $\widetilde{\lambda}_{\phi}^2$. This leads to the spectrum \subref{fig:singleReggeSwS}, in which the majority of states form a single linear trajectory. The few double states and outliers are most likely due to the restriction to a finite number of constraints and wavepackets. Pushing the cutoff (grey dashed line) higher leads to a drop in the $\widetilde{\lambda}_{\phi}^2$ bound.
        Both plots were generated with $k_{\text{max}}=17$ and $n_{\text{max}}=15$.}
        \label{fig:singleReggeSpectra}
\end{figure}

We now turn to the amplitude that extremizes the scalar coupling (green points labelled ``B'' in figure \ref{fig:CScolor}). We can first look at the spectrum that \texttt{SDPB} outputs, see figure \ref{fig:singleReggeSnS}. This spectrum describes a solution to the truncated bootstrap problem that, in principle, should approximate the extremal amplitude saturating the bound. However, one should interpret this solution with care, as it often appears to be the case that \texttt{SDPB} spectra for positivity bounds are populated by \textit{spurious} poles that are not part of the actual extremal solution \cite{Caron-Huot:2021rmr,Albert:2022oes,Albert:2023seb}.
Such states are often at or near the cutoff and, in general, increasing the number of null constraints and/or wavepackets does not help to tell them apart, as this increases both the number of physical and spurious states found by \texttt{SDPB}.
%In general, the number of states, both physical and spurious, found by \texttt{SDPB} increases with the number of null constraints and wavepackets.

The first thing we notice from figure \ref{fig:singleReggeSnS} is the beginning of a \textit{linear} Regge trajectory continuing the line from the graviton to the scalar at $m_\phi$ \textit{with the same slope}. This matches the graviton Regge trajectory that we would expect from string theory!
There are, however, many additional states to the left of this line, which is the opposite side compared to where string theory has its states. We expect these states to be unphysical and can test this by making further spectral assumptions. By not enforcing positivity of the sum rules within small areas of the parameter space (at low $J$ and $m^2$), we can prevent \texttt{SDPB} from finding solutions with states in these areas. We can therefore incrementally exclude more and more of the states above the Regge trajectory. If the bound on the scalar coupling does not change after these modifications, then states that were excluded are deemed unphysical. If, on the other hand, we exclude too much, the bound will drop.

Figure \ref{fig:singleReggeSwS} shows the result of this analysis: the necessary states lie on (or below) the linear Regge trajectory. 
Deviations from the exact line as well as the duplicates of states with the same spin and close in mass are presumably again a result of the (necessary) truncations for the numerical implementation of the bootstrap problem.
In fact, we have explored more stringent spectral assumptions excluding spin by spin ---for the first low-lying ones--- any states beyond the linear trajectory.\footnote{As a technical comment, we would like to point out that when using smeared sum rules (as in the current case) solutions are numerically much more unstable to spectral assumptions, compared to studies using only forward-limit sum rules. For these particular spectral assumptions, for example, one must allow for small boxes around the linear Regge trajectory, rather than single points on it.} These remained consistent with our extremal solution. From this, we conjecture that the ``physical'' amplitude that maximizes the scalar coupling only has states that lie on \textit{a single (linear) Regge trajectory}. That is, with no subleading (or daughter) Regge trajectories, in stark contrast with string theory.

This result is analogous in many ways to the recent finding in \cite{Albert:2023seb} of an extremal solution with a single (curved) Regge trajectory in the context of large $N$ pion scattering. That extremal solution was found there by extremizing the on-shell coupling of a spin-two state in the rho meson Regge trajectory. The intuition being that a spin-two exchange has really bad Regge behavior, and it necessitates a full Regge trajectory to UV complete it. The same is true here. A ``spin-zero'' exchange (really, a minimal massive supermultiplet) grows as $O(s^0)$ in the Regge limit, which is two steps worse compared to the assumed Regge behavior \eqref{eq:Regge}. As a result, it needs a spin-two state to kick in at a finite mass to UV complete it, which in turn needs a spin-four state, etc, furnishing a whole Regge trajectory.

What is surprising in these solutions is the apparent absence of daughter Regge trajectories. Under general assumptions about crossing symmetry and unitarity (not unlike the ones discussed here), and using standard techniques from Regge theory, it was proven in \cite{Eckner:2024pqt} that meromorphic and reggeizing amplitudes cannot be made of any finite number of Regge trajectories. This proof was for color-ordered amplitudes, but it is expected to generalize to the current case. So the question is \textit{where are the daughters?} A plausible scenario is that these are limiting amplitudes from continuous families of healthy amplitudes with an infinite number of Regge trajectories. If there is a consistent limit where all daughters are parametrically sent to infinity, or where their couplings vanish, the bootstrap will not be able to detect them and it will return only the limiting amplitude; much like how pure supergravity appeared as a limit in figure \ref{fig:CSg0g2}. Alternatively, there may exist an amplitude that honestly satisfies all of our assumptions; in that case, one of the assumptions~\cite{Eckner:2024pqt}
must be violated, possibly the absence
of Regge cuts.

\subsubsection*{Point C}\label{sec:shiftedRegge}
Finally, we would like to perform a similar analysis for the remaining amplitude (orange points labelled ``C'' in figure \ref{fig:CScolor}). 
Since its scalar coupling vanishes (i.e.\ $\widetilde \lambda_\phi^2\to 0$), one might expect it to be a simple rescaling of the amplitude at Point B. But, as was discussed previously, we observe that it is not. Its appearance in figures \ref{fig:CSg0gP} and \ref{fig:CSg0g2gP} for a small (but non-zero) value of $\widetilde \lambda_\phi^2$ makes manifest that this amplitude arises as yet another limit of a family of healthy amplitudes where $\widetilde \lambda_\phi^2\to 0$ but other couplings remain finite.

\begin{figure}[ht]
\centering
\includegraphics[width=0.8\textwidth]{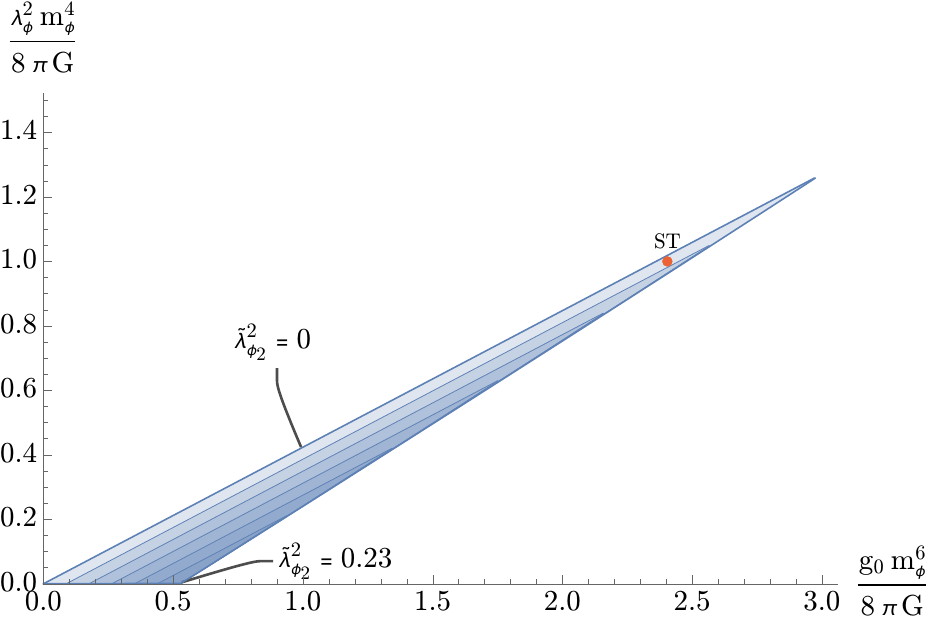}
\caption{Exclusion plot in $\widetilde{g}_0$ and $\widetilde{\lambda}_{\phi}^2$ as in figure \ref{fig:CSg0gP} but with the additional inclusion of the coupling $\widetilde{\lambda}_{\phi}^2$ of a second scalar with mass $2m_{\phi}^2$. This coupling is zero along the upper bound of the exclusion plot and increases linearly in the direction perpendicular to this line. Its maximum is reached at the lower extremal point. The red point marks the Virasoro-Shapiro amplitude \eqref{eq:MST}. The coupling of its scalar at mass $8/\alpha'$ matches that of the numerics. This plot was generated with $k_{\text{max}}=n_{\text{max}}=15$.}
\label{fig:CSsecondscalar}
\end{figure}

We start indirectly by turning our eye to states at mass $2m_{\phi}^2$. We can impose a second scalar exchange with coupling $\lambda_{\phi_2}$ and mass $m_{\phi_2}^2=2m_{\phi}^2$ and then regenerate the exclusion plot in $\widetilde{g}_0$-$\widetilde{\lambda}_{\phi}^2$ from figure \ref{fig:CSg0gP}. We find that lines of constant $\widetilde{\lambda}_{\phi_2}^2=\frac{m_{\phi}^4}{8\pi G}\lambda_{\phi_2}^2$ are parallel to the upper line that connects the pure supergrity amplitude (Point A) and the single Regge trajectory amplitude (Point B), see figure \ref{fig:CSsecondscalar}. 
Along this upper bound the coupling of the second scalar is zero, as was expected since the amplitudes at either end have no scalar at $2m_{\phi}^2$. The scalar coupling increases linearly in the perpendicular direction and is maximal at our point in question (Point C):
\begin{equation}\label{eq:SecondScalarBound}
    0\leq \widetilde{\lambda}_{\phi_2}^2\leq 0.231 \quad\quad (n_{\text{max}}=15,\,k_{\text{max}}=17).\hspace{-4cm}
\end{equation}
We have also looked at the intermediate value for which it crosses the Virasoro-Shapiro point (red dot in figure \ref{fig:CSsecondscalar}) and found exact agreement with the string theory result.

\begin{figure}[ht]
\centering
\includegraphics[width=0.6\textwidth]{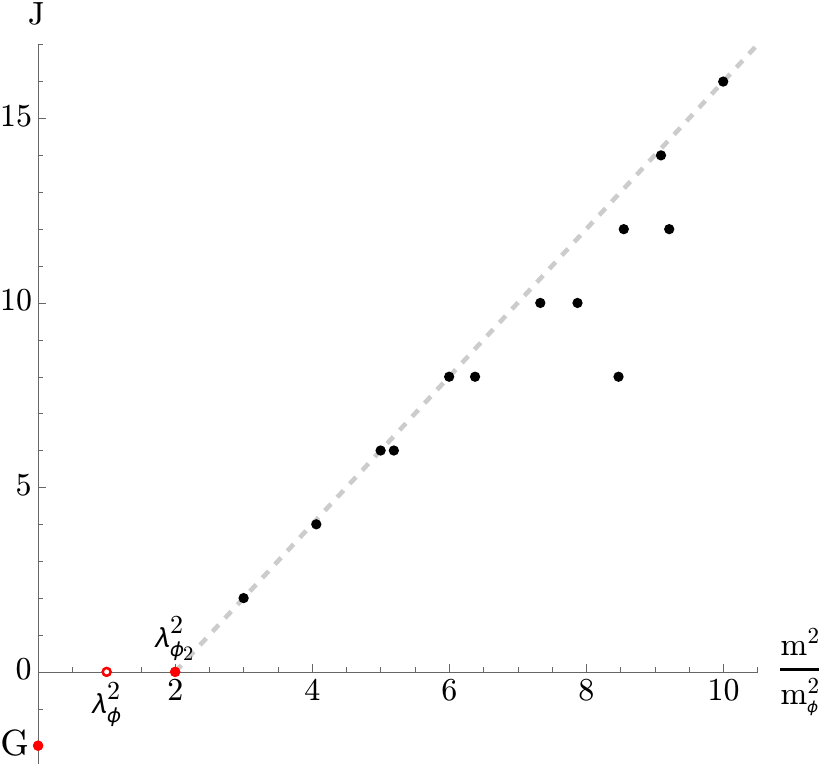}
\caption{Spectrum output from \texttt{SDPB} when maximizing the second scalar coupling, $\widetilde{\lambda}_{\phi_2}^2$, at mass $m_{\phi_2}^2=2m_{\phi}^2$ (i.e. Point C). The red ring denotes the first scalar, enforced with very small coupling $\widetilde{\lambda}_{\phi}^2=0.001$. The solid red dots denote the graviton and the second scalar. The grey dashed line shows the spin-dependent cutoff. (For spins $J>30$ the cutoff $M^2=2m_{\phi}^2$ is used.) Taking a higher cutoff (for low spins) decreases the upper bound on $\widetilde{\lambda}_{\phi_2}^2$. The majority of states form a single linear trajectory with a few double states and outliers which are expected to be due from the restriction to a finite number of constraints and wavepackets. From this we conjecture that the spectrum of this amplitude (which maximizes the coupling of the second scalar) is that of a single Regge trajectory but shifted. This plot was generated with $k_{\text{max}}=17$ and $n_{\text{max}}=15$.}
\label{fig:CSshifted}
\end{figure}

We can now proceed by looking at the \texttt{SDPB} spectrum at the point that maximizes the coupling of the mass $2m_{\phi}^2$ scalar. We find that one can again make spectral assumptions that exclude states in an upper triangle area, see figure \ref{fig:CSshifted}. The extremal amplitude seems to again only consist of a single Regge trajectory, this time shifted to cut the $J=0$ axis at $2m_{\phi}^2$. Notice that this is quite different from the spectrum that we would get by rescaling figure \ref{fig:singleReggeSwS} by $m_\phi^2 \to 2m_\phi^2$ so that the first resonance is at mass $2m_\phi^2$. That procedure would reduce the slope by half, whereas in figure \ref{fig:CSshifted} the trajectory is shifted but preserves the same slope.

We can now finally develop a qualitative understanding of the full geometry of the $\widetilde{g}_0$-$\widetilde{g}_2$ 
%understand the full $\widetilde{g}_0$-$\widetilde{g}_2$
plot from section \ref{sec:CSeftbounds}, i.e.\ figure \ref{fig:CSg0g2}. As a recap, the linear upper bound is saturated by (some UV-completion of) the tower of states amplitude \eqref{eq:Mtower}, until it flattens out and it (presumably) becomes a linear combination of this amplitude with the curious extremal solution with a single linear Reggge trajectory (Point B). The lower bound continuously connects the latter to the pure supergravity amplitude (point A); going through Point C.
We conjecture that this bound is saturated by a continuous family of amplitudes that, alongside the graviton pole, each has a single linear Regge trajectory (starting at a scalar exchange) such that one can freely dial the intercept while keeping the slope unchanged.
When the scalar is at the cutoff $M^2$, we would get the single Regge trajectory amplitude (recall figure \ref{fig:singleReggeSwS}) that we denoted by Point B in figure \ref{fig:CScolor}. Increasing the mass of the scalar exchange (and by this lowering the intercept of the Regge trajectory) would bring us downwards along the lower bound continuously. When the scalar mass hits $2M^2$, we would find the shifted single Regge trajectory (Point C) that we explored in this section. Further increasing the mass would push all states up higher and higher until they reach infinity (Point A). This picture would explain why only the part of the lower bound below Point C remained allowed in figure \ref{fig:CSg0g2gP}, where the cutoff was doubled with respect to figure \ref{fig:CSg0g2}.

\subsection{Analytical exploration}
Needless to say, a definitive understanding of the significance of the bounds only becomes available if one can find explicit \textit{analytical} amplitudes that saturate them. Exploring how much can be \textit{ruled in} by explicit solutions to the bootstrap is therefore an insightful game to play. We already saw in section \ref{sec:CSeftbounds} that linear combinations of string-theory amplitudes $M_{\text{ST}}(s,u)$ (defined in \eqref{eq:MST}) rule in a whole sliver which is well within our bounds. The question is \textit{can we get past this sliver?} Note that at this stage this does not yet require fully consistent UV completions of supergravity but merely four-point amplitudes compatible with all of our assumptions. 

\begin{figure}[h]
     \centering
     \begin{subfigure}[b]{0.495\textwidth}
         \centering
         \includegraphics[width=\textwidth]{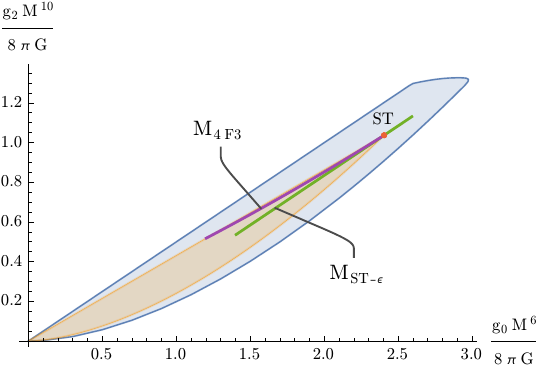}
         \caption{$\widetilde{g}_2$ vs. $\widetilde{g}_0$}
         \label{fig:Analytic(a)}
     \end{subfigure}
     \hfill
     \begin{subfigure}[b]{0.495\textwidth}
         \centering
         \includegraphics[width=\textwidth]{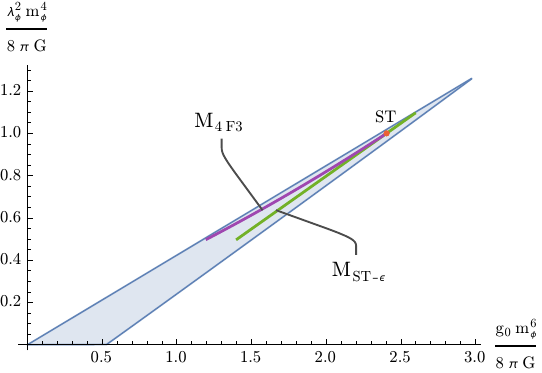}
         \caption{$\widetilde{\lambda}_\phi^2$ vs. $\widetilde{g}_0$}
         \label{fig:Analytic(b)}
     \end{subfigure}
        \caption{Location of two analytical amplitudes in the exclusion plots of \subref{fig:Analytic(a)} figure \ref{fig:CSg0g2} and \subref{fig:Analytic(b)} figure \ref{fig:CSg0gP}. The green line is spanned by the $M_{\text{ST}-\epsilon}(s,u)$ amplitude from \eqref{eq:MNima} for $\epsilon \in [-3/16,1]$. The purple curve corresponds to $M_{{}_4F_3}(s,u)$ for $x\in [0,1]$. Figure \subref{fig:Analytic(a)} shows these amplitudes for the cutoff $M^2\alpha'/4=1$. We can further rule in a whole sliver for every point ---similar to the (orange) one from the Virasoro-Shapiro amplitude--- by considering linear combinations of varying $\alpha'$.
        }
        \label{fig:analyticExp}
\end{figure}

One natural playground consists of amplitudes with the same spectrum as $M_{\text{ST}}(s,u)$ but with different on-shell couplings (i.e~residues). This can be achieved by shifting $M_{\text{ST}}(s,u)$ by Virasoro-Shapiro-like terms with differing arguments, often dubbed ``satellite terms'' in the literature. For example, in \cite{Arkani-Hamed:2020blm} they introduced the following deformed amplitude:
\begin{equation}\label{eq:MNima}
    M_{\text{ST}-\epsilon}(s,u) \equiv -\frac{\Gamma\left(\hbox{$-\frac{\alpha' s}{4}$}\right) \Gamma\left(\hbox{$-\frac{\alpha' t}{4}$}\right) \Gamma\left(\hbox{$-\frac{\alpha' u}{4}$}\right)}{\Gamma\left(\hbox{$1+\frac{\alpha' s}{4}$}\right) \Gamma\left(\hbox{$1+\frac{\alpha' t}{4}$}\right) \Gamma\left(\hbox{$1+\frac{\alpha' u}{4}$}\right)}-\epsilon\, \frac{\Gamma\left(\hbox{$1-\frac{\alpha' s}{4}$}\right) \Gamma\left(\hbox{$1-\frac{\alpha' t}{4}$}\right) \Gamma\left(\hbox{$1-\frac{\alpha' u}{4}$}\right)}{\Gamma\left(\hbox{$2+\frac{\alpha' s}{4}$}\right) \Gamma\left(\hbox{$2+\frac{\alpha' t}{4}$}\right) \Gamma\left(\hbox{$2+\frac{\alpha' u}{4}$}\right)}\,.
\end{equation}
For any $\epsilon$, this amplitude remains meromorphic (with the same spectrum as $M_{\text{ST}}$), crossing symmetric, and it enjoys strict spin-minus-two Regge behavior \eqref{eq:Regge} (as both terms grow as $\sim s^{-2+\alpha' u/4}$ in the Regge limit). It was shown in \cite{Arkani-Hamed:2020blm} that this amplitude is furthermore unitary in the range $\epsilon \in [0,1]$, but one can check numerically up to very high level that the residues remain positive in the extended range $\epsilon\in [-3/16,1]$. The new lower bound of this range occurs when the on-shell coupling of the $J=0$ state at mass $m^2=8/\alpha'$ vanishes. This family of amplitudes for varying $\epsilon$ traces the green lines in figure \ref{fig:analyticExp}. By further dialing $\alpha'$ we can cover a whole power-law sliver in the $\widetilde g_0-\widetilde g_2$ plane which, for $\epsilon<0$, extends beyond the area ruled in by string theory (i.e.\ the orange area in figure \ref{fig:Analytic(a)}). We have also checked that the coupling of the scalar at mass $8/\alpha'$ decreases along the green line in figure \ref{fig:Analytic(b)} in exact agreement with figure \ref{fig:CSsecondscalar}, reaching 0 when $\epsilon \to -3/16$; where the green line meets the boundary.

More generally, one can consider shifts of the form
\begin{equation}\label{eq:ansatz}
\sum_{\begin{smallmatrix}c_s,c_t,c_u\\ d_s,d_t,d_u\end{smallmatrix}} a_{(c_s,c_t,c_u),(d_s,d_t,d_u)}\frac{\Gamma\left(\hbox{$c_s-\frac{\alpha' s}{4}$}\right) \Gamma\left(\hbox{$c_t-\frac{\alpha' t}{4}$}\right) \Gamma\left(\hbox{$c_u-\frac{\alpha' u}{4}$}\right)}{\Gamma\left(\hbox{$d_s-\frac{\alpha' s}{4}$}\right) \Gamma\left(\hbox{$d_t-\frac{\alpha' t}{4}$}\right) \Gamma\left(\hbox{$d_u-\frac{\alpha' u}{4}$}\right)}\,,\quad (c_i,d_i\in \bZ_{+})
\end{equation}
where the coefficients $a_{(c_s,c_t,c_u),(d_s,d_t,d_u)}$ are fully symmetric in each set of indices and we restrict to terms with $(c_s+c_t)-(d_s+d_t)\leq -2$ (and similar for other channels) so that they do not spoil the Regge behavior. We could even consider integer rescalings of the slope, $\alpha'/4\to \alpha'/4k\; (k\in \bZ_+)$, so that the poles are separated by integer multiples of $4k/\alpha'$. Like in the case above, all these deformations leave the spectrum unchanged but modify the residues. Crossing symmetry and the Regge behavior are also built in (although one ought to be careful when considering infinite sums), so the constraining property is unitarity. In \cite{Haring:2023zwu}, they set up a primal bootstrap approach to systematically rule in allowed values for the coefficients $a_{c,d}$ in the case of MHV graviton scattering amplitudes in four dimensions. It would be interesting to carry out that program in the current case of ten-dimensional supergravity with maximal supersymmetry.

We have not pursued this direction systematically, but we did stumble upon a deformation of the form \eqref{eq:ansatz} which seems to check all the boxes. Namely,
\begin{align}\label{eq:4F3}
    M_{{}_4F_3}(s,u) \equiv&\, -\sum_{n=0}^\infty \frac{\Gamma\left(\hbox{$n-\frac{\alpha' s}{4}$}\right) \Gamma\left(\hbox{$n-\frac{\alpha' t}{4}$}\right) \Gamma\left(\hbox{$n-\frac{\alpha' u}{4}$}\right)}{\Gamma\left(\hbox{$1+n+\frac{\alpha' s}{4}$}\right) \Gamma\left(\hbox{$1+n+\frac{\alpha' t}{4}$}\right) \Gamma\left(\hbox{$1+n+\frac{\alpha' u}{4}$}\right)}\frac{(x)_n}{n!}\\
    =&\, -\frac{\Gamma\left(\hbox{$-\frac{\alpha' s}{4}$}\right) \Gamma\left(\hbox{$-\frac{\alpha' t}{4}$}\right) \Gamma\left(\hbox{$-\frac{\alpha' u}{4}$}\right)}{\Gamma\left(\hbox{$1+\frac{\alpha' s}{4}$}\right) \Gamma\left(\hbox{$1+\frac{\alpha' t}{4}$}\right) \Gamma\left(\hbox{$1+\frac{\alpha' u}{4}$}\right)} {}_4F_3\left[\begin{matrix}
        -\frac{\alpha's}{4},\, -\frac{\alpha't}{4},\,-\frac{\alpha'u}{4},\,x\\
        1+\frac{\alpha's}{4},1+\frac{\alpha't}{4},1+\frac{\alpha'u}{4}
    \end{matrix}\,;1\right]\,.\nonumber
\end{align}
This amplitude was inspired by \cite{Cheung:2023adk}, where they introduced a similar-looking deformation of the Veneziano amplitude. In the latter form of this amplitude, we see that all the poles come from the Virasoro-Shapiro factor and the hypergeometric function simply replaces the residues by different polynomials of the same degree. We have checked numerically that the residues of poles up to very high level remain positive in the range $x\in [0,1]$. This amplitude traces the purple curves in figure \ref{fig:analyticExp} as we tune $x$ in this range. While in figure~\ref{fig:Analytic(a)} the whole curve is contained in the sliver already ruled in by string theory, we see that (through linear combinations with itself and/or with $M_{\text{ST}-\epsilon}(s,u)$) it significantly extends the ruled in area in figure \ref{fig:Analytic(b)}.

This example highlights the freedom that might be encoded in deformations of the Virasoro-Shapiro amplitude of the form \eqref{eq:ansatz}. One could potentially rule in the whole allowed region in the various exclusion plots by only considering such deformations. In particular, we just discussed in section \ref{sec:numAnalysis} how the numerics point towards a curious bootstrap solution with a \textit{single Regge trajectory}. We note that the ansatz \eqref{eq:ansatz} has (more than) enough free parameters to cancel the subleading Regge trajectories of $M_{\text{ST}}(s,u)$ state by state. So, in principle, one could conceive an infinite sum with coefficients engineered to leave only the leading trajectory. The question is whether such a sum would converge (and not spoil other properties along the way).\footnote{The way this infinite sum could escape the no-go theorem of \cite{Eckner:2024pqt} is by creating a Regge cut or some other singularities in the complex $J$ plane. It was demonstrated in \cite{Veneziano:2017cks} (in the context of large $N$ pion scattering) that a Regge cut is produced when considering a certain infinite sum of Lovelace-Shapiro-like amplitudes. One expects similar singularities may be created by the infinite sum here.} This question remains open. But either way, we do not necessarily expect these amplitudes to be fully consistent theories. Rather, they are more likely to be artifacts of four-point supergravion scattering. Upgrading to constraints from higher-point bootstrap, or the mixed scattering system with excited states in the external legs should help us decide.

\section{Setup for supergluon scattering
} \label{sec:openStrings}
Let us change gears now and return to the scattering of $\mathcal N=(1,0)$ super Yang-Mills (SYM) multiplets. We saw in section \ref{sec:N=(1,0)} that ---like in the case of supergraviton scattering with maximal supersymmetry--- all the four-point amplitudes of this multiplet are determined in terms of a single auxiliary scalar amplitude $M(s,u)$. This would cease to be the case if we included the $(1,0)$ supergraviton, which is why we will restrict to pure gauge interactions in this section, leaving the extension to gravity as an interesting future direction.
More precisely, we will work in the limit of weak gauge coupling, $g_{\rm YM}\to 0$, where we have already turned off the gravitational coupling $G = 0$. Note that this restriction excludes heterotic string theories, in which the gauge and gravitational couplings are both proportional to the \textit{closed string coupling} $g_{\rm YM}^2\sim G\sim g_{s}^2$. It is compatible, however, with the tree-level limit of Type~I string theory ---a theory of unoriented \textit{open strings} with $G\sim g_s^2$ but $g_{\rm YM}^2\sim g_{s}$.

Compared to the above analysis for supergravitons, the setup here will be less canonical. For reasons that will become apparent as we go along, in order to make progress we will need to make more stringent assumptions which, as far as we can tell, can only be motivated in a string theory context. We will therefore have in mind a theory of open strings, following the standard kinematics, but with the spectrum and other dynamical data unfixed. The question we will thus be addressing is whether Type~I string theory is the only theory of open strings which UV-completes super Yang-Mills theory at tree level.

In section \ref{sec:N=(1,0)} we omitted the color indices of the SYM multiplet, we will introduce them now. Massless vectors describe gauge fields of some Lie group $G$. As such, they (and their superpartners) are valued in the adjoint representation of $G$. Due to a famous  anomaly, the ten-dimensional SYM theory is only consistent for $G=SO(32)$ or $E_8 \times E_8$~\cite{Green:1984sg}. We will restrict to the former, which is known to be realized by Type~I string theory,
although the dependence of our results on the gauge group will be minimal.
In this case, each gluon is associated to a generator $T_a$ of $SO(32)$,\footnote{We take these generators in the fundamental representation, in a basis where they are purely imaginary and antisymmetric, $T_a^\intercal = -T_a$. We normalize them by $\tr{T_a T_b} = \frac{1}{2}\delta_{ab}$, and our conventions for the (real) structure constants are $f_{abc} \equiv \frac{2}{i}\tr{T_a[T_b,T_c]}$. The $d$-symbols $d_{abc} \equiv 2\tr{T_a\{T_b,T_c\}}$ vanish by antisymmetry.} introduced via Chan-Paton factors at the ends of the corresponding open string. At tree level, only worldsheets with the topology of a disk (with a color trace running along the boundary) contribute, so the four-gluon amplitude takes the form
\begin{equation}\label{eq:Tabcd}
    {\mathcal T}_{abcd}  = \tr{T_aT_bT_cT_d} \mathcal{F}^4 M(s, t)
    + \tr{T_aT_bT_dT_c} \mathcal{F}^4 M(s, u)
    + \tr{T_aT_cT_bT_d} \mathcal{F}^4 M(t, u)\,.
\end{equation}
Here the factor $\mathcal F^4$ (defined in \eqref{eq:F4def}) takes care of the polarizations of the gluons, and other amplitudes in the supermultiplet can be obtained by acting with SUSY transformations on it.
It follows from Bose symmetry of $\mathcal T$ (and the full crossing symmetry of $\mathcal F$) that $M(s,u)$ is only $s\leftrightarrow u$ symmetric.

\subsection{Analyticity, unitarity, Regge behavior}
Once again, in the tree-level limit, $M(s,u)$ is analytic throughout the complex plane except for singularities due to the physical exchanges of states in the spectrum. There is a massless pole corresponding to the self-interactions of the SYM multiplet, and a collection of massive singularities ---due to the excited states--- whose precise configuration we do not know.
We will restrict to UV completions of SYM theory whose excited states are \textit{all valued in the adjoint of $G$}.
This is automatic in open string theory, where excited states are also open-string states with the same Chan-Paton factors as the gluon.
We leave it as an interesting future direction to explore more exotic UV completions with excited states in general representations of $G$.

To evaluate the consequences of unitarity, we first need to expand the full four-gluon amplitude \eqref{eq:Tabcd} in partial waves. This involves an expansion in little group representations (or ``spins'') as well as representations $\mathcal R$ of $G$ that can be exchanged in a given channel. The color expansion trivializes when assuming that all excited states are valued in $\mathcal R=\text{adj}$ of $G=SO(32)$. In this case, the only invariant three-point tensor is $f_{abc}$, so we have that physical exchanges can come only with color structures of the form $f\indices{_{ab}^e}f_{cde}$, and its crossed versions. Using the identity
\begin{equation}
    f\indices{_{ab}^e}f_{cde} = -2\tr{[T_a,T_b][T_c,T_d]}\,,
\end{equation}
it is straightforward to see that each of the terms in \eqref{eq:Tabcd} receives contributions from only two channels. As a result, the color-ordered amplitude $\mathcal{F}^4 M(s, u)$ will be \textit{free from singularities in the $t$ channel}.

Unitarity for the color-ordered amplitude $\mathcal{F}^4 M(s, u)$ in $D$ dimensions was discussed in \cite{Arkani-Hamed:2022gsa}. It was shown there that the polarization prefactor admits an $s$-channel expansion $\mathcal F^4 = \langle 1,2|\mathcal O |3,4\rangle$, in terms of a diagonal operator representing the exchange of states with spin zero, spin two and a three-form. Schematically,
\begin{equation}\label{eq:F4Decomp}
\mathcal{O} = \left(\frac{9}{D{-}1}{-}1\right) \big| \cdot \big\rangle \big\langle \cdot\big| 
+ \big|\,{\tiny\ytableausetup{centertableaux}\ydiagram{2}}\, \big\rangle \big\langle \,{\tiny \ytableausetup{centertableaux} \ydiagram{2}}\,\big| 
+ \big|\,{\tiny \ytableausetup{centertableaux} \ydiagram{1,0+1,0+1}}\, \big\rangle \big\langle \,{\tiny \ytableausetup{centertableaux} \ydiagram{1,0+1,0+1}}\,\big| \,.
\end{equation}
More generally, in the intermediate states we can have these $SO(D-1)$ representations tensored with a symmetric spin-$J$ irrep,
\begin{equation*}
    \left( \frac{9}{D{-}1} {-}1\right)\big| \cdot\otimes J\big\rangle\big\langle \cdot\otimes J\big|
    + \big|\,{\tiny\ytableausetup{centertableaux} \ydiagram{2}} \otimes J\big\rangle \big\langle \,{\tiny \ytableausetup{centertableaux} \ydiagram{2}}\otimes J\big|
    + \big|\,{\tiny\ytableausetup{centertableaux} \ydiagram{1,0+1,0+1}} \otimes J\big\rangle \big\langle \,{\tiny \ytableausetup{centertableaux} \ydiagram{1,0+1,0+1}} \otimes J\big| \,. 
\end{equation*}
Factoring out the polarization factor $\mathcal F^4$, one then finds that the partial wave expansion for the full amplitude \eqref{eq:Tabcd} reduces to the familiar scalar partial wave expansion for the reduced amplitude $M(s,u)$ \cite{Arkani-Hamed:2022gsa}. Namely,
\begin{equation}
    {\rm Im} \,  M(s,u)  = s^{\frac{4-D}{2}} \sum_{J} n^{(D)}_J \rho_J(s)\, \legP_J\left( {1+\frac{2u}{s}} \right) \,,
\end{equation}
where spins can now be even and odd. Unitarity is now the usual statement that $\rho_J(s) \geq 0$ in the physical region.

As far as the Regge behavior of $M(s,u)$ goes, we will assume that it is strictly better than that of the gluon exchange.
The reason is that, in string theory-like amplitudes, we expect the gluon to reggeize. That is, we expect it to join the excited states in a leading Regge trajectory which locally looks like $\alpha(u)\approx \alpha_0 + \alpha' u + O(s^2)$, with $\alpha_0$ the spin of the gluon. Based on general assumptions of unitarity and causality, Regge theory then predicts that reggeizing amplitudes grow as $s^{\alpha(u)}$ in the Regge limit \cite{Gribov:2003nw}.
Since the gluon has spin one, its exchange grows roughly as $\mathcal T\sim s$. Removing the polarization factor $\mathcal{F}^4$ reduces it by two powers of $s$, leaving the Regge behavior
\begin{equation}\label{eq:ReggeOpen}
    \lim_{|s| \to \infty} s M(s, u) = 0 \, ,  \quad u < 0 \,.
\end{equation}
It would be remarkable if one could show that this is the only possibility for any consistent UV-completion of SYM, without relying on string theory arguments. This assumption will be essential later to capture the gauge coupling $g_{\text{YM}}$ in dispersion relations and thus couple it to the excited states.

\subsection{Low energies: Corrections to super Yang-Mills}\label{sec:lowESYM}
From this point onwards, the discussion is very parallel to the one in section \ref{sec:closedStringsSetup} for supergravitons. At low energies, the scattering of $\mathcal N=(1,0)$ gluon supermultiplets is dictated by super Yang-Mills theory. The corresponding amplitude reads
\begin{equation}\label{eq:MSYM}
    M_{\text{SYM}}(s,u) = -\frac{g_{\text{YM}}^2}{su}\,,
\end{equation}
with $g_{\text{YM}}$ the Yang-Mills coupling. The pole describes either the exchange of a gluon or of a gaugino depending on which component of the superamplitude we are considering. This amplitude fails to satisfy the Regge behavior \eqref{eq:ReggeOpen}, so the full theory will necessitate massive states that UV-complete this exchange. Again, these can be parametrized by a collection of on-shell data $\{m_i^2,J_i,\lambda_{ijk}\}$.

By integrating out all these excited states we obtain a low-energy effective amplitude in terms of a set of unknown Wilson coefficients $\{g_i\}$, valid up to a cutoff $M^2$, below the mass of the first excited state. In this case, the expansion reads
\begin{equation}\label{eq:MlowOpen}
    M_\text{low}(s,u) = -\frac{g_{\text{YM}}^2}{su} + g_0 + g_1 (s + u) + g_2(s^2 + u^2) + g_2' s u + \ldots\,,
\end{equation}
since the color-ordered amplitude is only $s\leftrightarrow u$ symmetric. 
Except for the gluon pole, this expansion is analogous to the one discussed in \cite{Albert:2022oes} for the low-energy scattering of pions, with the difference of the $g_0$ term, which is not prevented by the Adler zero here. Another difference with that case is that the improved Regge behavior \eqref{eq:ReggeOpen} will allow us to normalize couplings by $g_{\text{YM}}^2$ rather than by the effective coupling $g_1$.

Following closely our analysis for closed strings, we will soon refine our assumptions to include the first massive state, which we will take to be in the smallest massive representation of the supersymmetry algebra. This is the only possibility when the gluon reggeizes, as the first massive state in its trajectory necessarily has effective spin $J_\text{eff}=0$.
For the current case of $\mathcal N=(1,0)$ supersymmetry, the minimal massive supermultiplet contains the $SO(9)$ representations
\begin{equation}\label{eq:openMassiveSupermultiplet}
    (\mathbf{44} + \mathbf{84} + \mathbf{128})\,.
\end{equation}
We recognize the spin two $\tiny\ytableausetup{centertableaux}\ydiagram{2}$ and the three-form $\tiny \ytableausetup{centertableaux} \ydiagram{1,0+1,0+1}$ from \eqref{eq:F4Decomp} with $D=10$, as well as a fermionic $\mathbf{128}$ ``spin-$3/2$'' state. In practice, integrating this massive state back into the EFT simply amounts to including an explicit ``scalar'' massive pole in the basic amplitude $M(s,u)$,
\begin{equation}\label{eq:MopenLowwPole}
    M_\text{low}^{\phi-\text{pole}}(s,u) = -\frac{g_{\text{YM}}^2}{su} + \lambda_{\phi}^2\left(\frac{1}{m_\phi^2 - s} + \frac{1}{m_\phi^2 - u}\right) + \text{analytic} \,.
\end{equation}

\subsection{The example: Type I string theory}\label{sec:TypeIstringthy}
The only theory of open strings known to UV-complete $SO(32)$ SYM theory in ten dimensions is Type I superstring theory, a theory of open and closed strings. Open strings form a consistent subsector at tree level.
The auxiliary $2 \to 2$ amplitude takes the following Veneziano-like form \cite{Green:1987sp},
\begin{equation}\label{eq:MopenST}
    M_{\text{ST}}(s,u) = -\frac{\Gamma(-\tfrac{\alpha'}{2} s) \Gamma(-\tfrac{\alpha'}{2} u)}{\Gamma(1+\tfrac{\alpha'}{2} t)}\,.
\end{equation}
It satisfies all of the assumptions discussed above, and it is further meromorphic with evenly distributed poles at masses $m^2 = 0, \frac{2}{\alpha'}, \frac{4}{\alpha'},...\,$. 
By expanding around $s,u\sim 0$ and comparing to \eqref{eq:MlowOpen} we extract the following low-energy coefficients:
\begin{gather}
    \text{ST:}\quad g_{\text{YM}}^2= \frac{4}{\alpha'^2}\,,
    \qquad  g_0 = \zeta(2) = 1.64493...\,,
    \qquad  g_1 = \zeta(3)\frac{\alpha'}{2} = \alpha' 0.601028...\,, \nonumber\\
     g_2 =  \zeta(4)\frac{\alpha'^2}{4} = \alpha'^2 0.270581...\,,
    \qquad  g_2'  = \frac{1}{4}\zeta(4)\frac{\alpha'^2}{4} = \alpha'^2 0.0676452... \,.
\end{gather}
The cutoff for this expansion can be pushed only up to the first massive pole in \eqref{eq:MopenST}, so $M^2\leq 2/\alpha'$.
At higher energies (namely $2/\alpha'\leq M^2 \leq 4/\alpha'$), the Type I amplitude \eqref{eq:MopenST} agrees with the refined EFT \eqref{eq:MopenLowwPole}, as the first massive open string mode furnishes exactly the supermultiplet \eqref{eq:openMassiveSupermultiplet}. By computing the residue, we determine the on-shell data
\begin{equation}
    \text{ST:}\qquad m_\phi^2 = 2/\alpha'\,, \quad J_\phi=0\,,\quad \lambda_\phi^2 = 2/\alpha'\,.
\end{equation}

\subsection{Positivity bounds from dispersion relations}\label{sec:posboundsOpen}
Dispersion relations follow the general recipe discussed above in section \ref{sec:closedStringsSetup} and references therein. The only subtlety for open strings is that there are now \textit{two} inequivalent sets of dispersion relations; those at fixed $u$ and those at fixed $t$. This was discussed in detail in~\cite{Albert:2022oes} in the context of pion scattering. After shrinking the contours in the usual manner, we obtain (for $u<0$),
\begin{subequations}\label{eq:disprelsOpen}
\begin{align}\label{eq:disprelsOpenA}
     \frac{1}{2 \pi i } \oint_{s\sim M^2} \frac{ds}{s}\,\frac{M_{\text{low}}(s, u)}{s^{k}} =&\,
    \avg{   \frac{ \legP_J(  1+\frac{2u}{m^2})}{m^{2k}}} \,,\\
    \frac{1}{2 \pi i } \oint_{s\sim M^2} \frac{ds}{s}\,\frac{M_{\text{low}}(s, -s-u)}{s^{k}} =&\, 
    \avg{ \left(\frac{1}{m^{2k}}+\frac{(-1)^k m^2}{(m^{2} + u )^{k+1}}\right)(-1)^J\legP_J(  1+\frac{2u}{m^2})}\,.
\end{align}
\end{subequations}
Both of these sets are valid for $k=-1,0,1,2...\,$. Of all these dispersion relations, only \eqref{eq:disprelsOpenA} for $k=-1$ is special, since it captures the gluon pole,
\begin{equation}
    -\frac{g_{\text{YM}}^2}{u} = \avg{   m^{2}\legP_J(  1+\tfrac{2u}{m^2})} \,.
\end{equation}
We will follow \cite{Caron-Huot:2021rmr} (like in section \ref{sec:posbounds} for supergravitons) and smear the pole against some wavepackets $f_k(p) = \{p^{3/2},p^{5/2},p^{7/2},... \}$ to access this sum rule. Namely,
\begin{equation}\label{eq:YMsmeared}
    g_{\text{YM}}^2 \int_0^{M} \hspace{-5pt}dp \, \frac{f_k(p)}{p^2} = \avg{\int_0^{M}\hspace{-5pt} dp \, f_k(p)\, m^2 \legP_J(1-\tfrac{2p^2}{m^2})}\,.
\end{equation}
The remaining dispersion relations in \eqref{eq:disprelsOpen} are standard and can be expanded in the forward limit $u\sim 0$ to extract sum rules. The ones that will be most useful to us are
\begin{equation}\label{eq:MSumRulesOpen}
    g_0 =  \avg{1}\,, \quad 
    g_1 =  \avg{\frac{1}{m^2}}\,, \quad
    g_2 =  \avg{\frac{1}{m^4}}\,, \quad
    g_2' = \avg{\frac{2\left(1-(-1)^J\right)}{m^4}}\,.
    %g_2' = \avg{\frac{2J(J+D-3)}{(D-2)}\frac{1}{m^4}}\,.
\end{equation}

There are in this case two infinite sets of null constraints, following from the two sets in \eqref{eq:disprelsOpen}. As explained in \cite{Albert:2023jtd}, an efficient way to compute them is to consider the double contour integral in \eqref{eq:NCint} separately for $F(s,u)=M(s,u)$ and $F(s,u)=M(-s-u,u)$. We obtain in this way the null constraints $\avg{\mathcal X_{n,\ell}}=0$ and $\avg{\mathcal Y_{n,\ell}}=0$, where
\begin{subequations}
\begin{align}
    \mathcal X_{n,\ell} =&\, \mathop{\mathrm{Res}}_{u = 0}
    \left[\frac{1}{u}\left(\frac{1}{m^{2(n-\ell)}u^\ell} - \frac{1}{u^{n-\ell} m^{2\ell}}\right)
    \legP_J(1+\tfrac{2u}{m^2})
    \right]\,.\\
    \mathcal Y_{n,\ell} =&\, \mathop{\mathrm{Res}}_{u = 0}\left[\frac{1}{u}
    \left(\frac{(-1)^{J+1}}{u^{n-\ell} m^{2\ell}}
    -\frac{m^2}{(-m^2-u)^{n-\ell+1} u^{\ell}} + \frac{(-1)^Jm^2}{u^{n-\ell} (-m^2-u)^{\ell+1}}\right)
    \legP_J(1+\tfrac{2u}{m^2})
    \right]\,.
\end{align}
\end{subequations}
An independent set is given by $n=1,2,...$, $\ell=0,1,...,\left[\frac{n-1}{2}\right]$ for $\mathcal X_{n,\ell}$ and $n=0,1,...$, $\ell=0,1,...,\left[\frac{n}{2}\right]$ for $\mathcal Y_{n,\ell}$. One can check that these formulas reproduce the null constraints from \cite{Albert:2022oes}.

Positivity bounds for (normalized ratios of) Wilson coefficients or on-shell couplings can then be derived through bootstrap equations entirely analogous to the ones discussed in preceding sections. For example, using $k_\text{max} = 27$ for the wavepackets and $n_\text{max} = 27$ for the null constraints in the analog of \eqref{eq:g0BootstrapEq} yields the optimal bound
\begin{equation}
    \widetilde g_0 \equiv \frac{g_0 M^{4}}{g_{\text{YM}}^2} \leq 2.59\,.
\end{equation}

\section{Results for supergluons}\label{sec:OSResults}
\subsection{EFT bounds}\label{sec:OSeftbounds}
\begin{figure}[ht]
\centering
\includegraphics[width=0.7\textwidth]{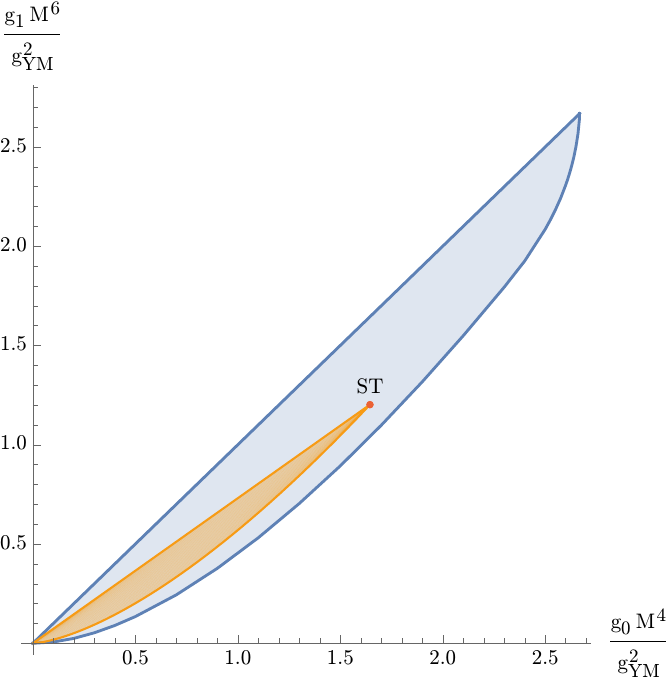}
\caption{Exclusion plot in the space of the first two EFT coefficients, $g_0$ and $g_1$, normalized by $g_{\text{YM}}^2$ and the cutoff $M$. The (blue) shaded region is the allowed space for amplitudes. It was generated numerically using $k_{\text{max}}=n_{\text{max}}=15$ (i.e. a total of 7 wavepackets and 136 null constraints). The red dot corresponds to the Type I amplitude \eqref{eq:MopenST} with $M^2\alpha'=2$. Varying $M^2\alpha'$ from 0 to 2 rules in the orange region.}
\label{fig:OSg0g1}
\end{figure}

We can now continue with a similar analysis to that of section \ref{sec:CSeftbounds},
for the current case of supergluon scattering. 
We begin by carving out the allowed region in the space of the lowest two Wilson coefficients, $\widetilde{g}_0 \equiv \frac{M^4}{g_{\text{YM}}^2}g_0$ and $\widetilde{g}_1 \equiv \frac{M^6}{g_{\text{YM}}^2}g_1$ defined in \eqref{eq:MlowOpen}, see figure \ref{fig:OSg0g1}.
Just as in figure \ref{fig:CSg0g2}, we find a linear upper bound (given here by $\widetilde{g}_1\leq \widetilde{g}_0$) which is exactly equal to the naive bound derived from disregarding all null constraints and just looking at the relevant sum rules \eqref{eq:MSumRulesOpen}, i.e.\ $\avg{M^2/m^2}\leq\avg{1}$.
In contrast with figure \ref{fig:CSg0g2}, though, the curved lower bound runs smoothly from the origin all the way to the linear upper bound, yielding no plateau.

Once again, the linear upper bound can only be saturated by an amplitude all of whose (finite mass) states sit exactly at the cutoff.
One such tower-of-states amplitude is given by~\cite{Berman:2023jys}
\begin{equation}\label{eq:MtowerOpen}
    M_\text{tower}(s,u) = -\frac{g_{\text{YM}^2}}{su} + \frac{\chi}{(M^2-s)(M^2-u)}\,,
\end{equation}
but this one (marginally) violates the assumed Regge behavior \eqref{eq:ReggeOpen}. The numerics indicate that there exists at least one family of healthy amplitudes (presumably with states at parametrically large mass) which limits to this one for a finite range of $\chi$. It would be interesting to pin down one such family explicitly.

\begin{figure}[ht]
\centering
\includegraphics[width=0.7\textwidth]{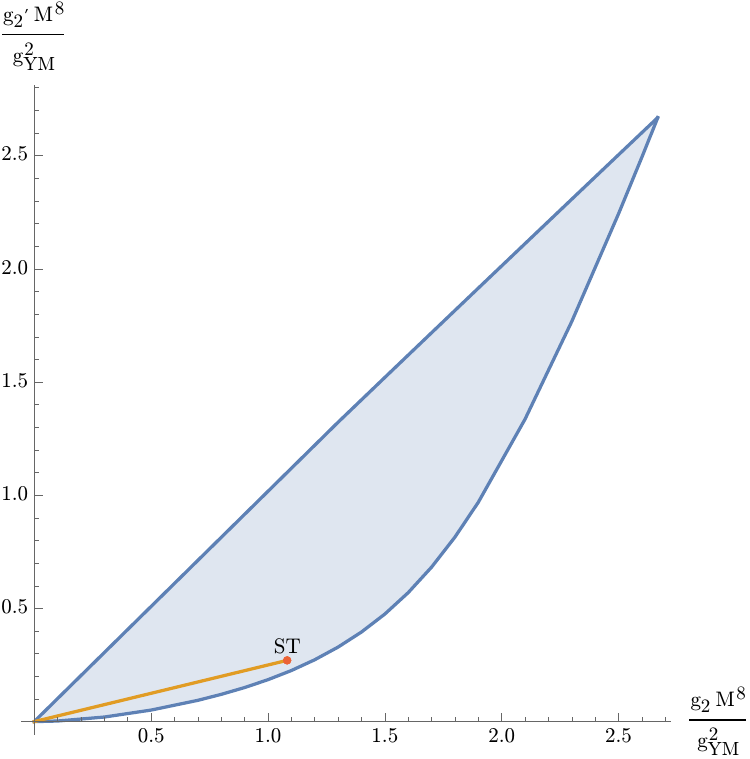}
\caption{Exclusion plot in the space of two higher EFT coefficients, $g_2$ and $g_2'$, normalized by $g_{\text{YM}}^2$ and the cutoff $M$. The (blue) shaded region is the allowed space for amplitudes. It was generated numerically using $k_{\text{max}}=n_{\text{max}}=15$ (i.e. a total of 7 wavepackets and 136 null constraints). The red dot corresponds to the Type I amplitude \eqref{eq:MopenST} with $M^2\alpha'=2$. Varying $M^2\alpha'$ from 0 to 2 rules in the orange line.}
\label{fig:OSg2g2p}
\end{figure}

We can also look at higher Wilson coefficients. Figure \ref{fig:OSg2g2p} shows the exclusion plot in the space of couplings $\widetilde{g}_2 \equiv  \frac{M^8}{g_{\text{YM}}^2}g_2$, $\widetilde{g}_2' \equiv \frac{M^8}{g_{\text{YM}}^2}g_2'$. 
The upper bound is also in this case given by a straight line, $\widetilde g_2'\leq \widetilde g_2$. From \eqref{eq:MSumRulesOpen} we learn that this bound is saturated when
\begin{equation}
    \avg{\frac{1}{m^4}}_{J=\text{even}} = 3\avg{\frac{1}{m^4}}_{J=\text{odd}}\,,
\end{equation}
but this condition does not look constraining enough to nail down the corresponding extremal amplitude. The lower bound again connects smoothly the origin with the upper right corner, featuring no special points along the boundary. This is qualitatively different from what was found in figure 5 of \cite{Berman:2023jys}, which studied the same Wilson coefficients (albeit normalized by $g_0$ rather than $g_\text{YM}^2$) in the context of $\mathcal N=4$ open-string scattering in four dimensions.\footnote{Our results can be compared at a qualitative level, despite the difference in dimensions, because positivity bounds tend to vary slowly with $D$, without significantly changing any features, see e.g.\ \cite{Caron-Huot:2021rmr}.} In that case, the lower bound went through a sharp corner with $\widetilde g_2'=0$, conjectured to correspond to an amplitude with a single scalar exchange,
\begin{equation}
    M_\text{scalar}(s,u) = -\frac{g_{\text{YM}}^2}{su} + \lambda^2\left(\frac{1}{M^2 - s} + \frac{1}{M^2 - u}\right) \,.
\end{equation}
Since \cite{Berman:2023jys} assumed a spin-zero Regge behavior, this amplitude was only borderline disallowed there, and such amplitudes can typically be ruled in by adding suitable infinitely-heavy states~\cite{Albert:2022oes}. Compared to our stronger Regge behavior assumption \eqref{eq:ReggeOpen}, however, this amplitude grows too fast by a full power ---it is therefore completely ruled out in figure~\ref{fig:OSg2g2p}.

In both exclusion plots the Type I amplitude \eqref{eq:MopenST} ---including the convex hull shaped by changing $\alpha'$--- sits within the allowed region.\footnote{In the $\widetilde{g}_2$-$\widetilde{g}_2'$ case (figure \ref{fig:OSg2g2p}), the convex hull is just a line since the mass dimensions of the two couplings are the same.} Only for $\alpha'\rightarrow 0$ does the ruled in region reach the boundary, providing a simple UV completion of the pure SYM amplitude \eqref{eq:MSYM}. In the $\widetilde{g}_2$-$\widetilde{g}_2'$ plot, the $M^2\alpha'=2$ point sits quite close to the boundary (see the red dot in figure \ref{fig:OSg2g2p}), but taking the limit of infinite wavepackets and null constraints will unlikely close the gap.
What additional constraints do we need to corner string theory?\footnote{It was demonstrated in \cite{Huang:2020nqy,Berman:2023jys,Chiang:2023quf} that the combination of positivity bounds with the so-called monodromy relations is constraining enough to corner string theory to very high precision. We do not incorporate these constraints here because we want to keep to a minimum
the input from a worldsheet description.}
We will proceed by refining our assumptions about the low-lying spectrum.

\subsection{Adding a scalar exchange}
Let us apply the same restriction on the spectrum as we did for supergravitons: the first massive resonance should consist of an isolated state of effective spin $J_\text{eff}=0$ (i.e.\ it is in the minimal massive supermultiplet \eqref{eq:openMassiveSupermultiplet}). This is done by considering the low-energy amplitude \eqref{eq:MopenLowwPole} (which depends explicitly on the mass $m_\phi$ and the coupling $\lambda_\phi^2$ of the scalar) or, equivalently, by shifting the high-energy spectral density as in \eqref{eq:rhoShift}. Like in section~\ref{sec:ScalarExch}, we will take $m_\phi$ to set the scale of the problem, and tune the cutoff $m_\phi^2 \leq M^2 <\infty$ after which new states may kick in.

\begin{figure}[ht]
\centering
\includegraphics[width=0.8\textwidth]{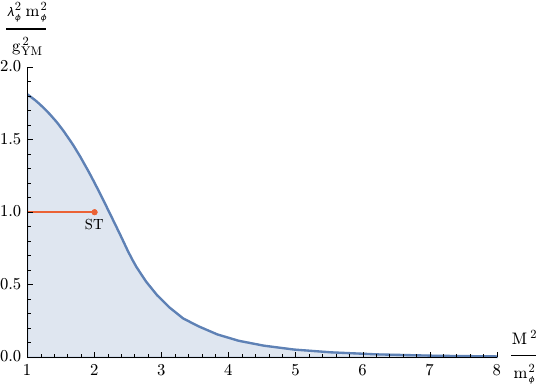}
\caption{Upper bound on the coupling of the first scalar, normalized by its mass $m_{\phi}^2$ and $g_{\text{YM}}^2$, as a function of the cutoff $M^2$. The shaded region is the allowed space for amplitudes. One example, the Veneziano amplitude with $M^2\alpha'=4$, is marked by the red dot; the red line is spanned by choosing an ``unnatural'' cutoff in $2/\alpha' \leq M^2 <4/\alpha'$. This plot was generated mostly with $k_{\text{max}}=13$ and $n_{\text{max}}=8$ (i.e. a total of 6 wavepackets and 45 null constraints). For cutoffs $M^2\geq 2m_{\phi}^2$ the number of null constraints was lowered to $n_{\text{max}}=6$ (28 null constraints).}
\label{fig:OSgPM1}
\end{figure}

Fixing the cutoff at $M^2=m_{\phi}^2$ we find the following bound on $\widetilde{\lambda}_{\phi}^2=\frac{m_{\phi}^2}{g_{\text{YM}}^2}\lambda_\phi^2$:
\begin{equation}
    0\leq \widetilde{\lambda}_{\phi}^2 \leq 1.81 \quad\quad (n_{\text{max}}=10,\,k_{\text{max}}=13).\hspace{-4cm}
\end{equation}
Increasing the cutoff leads to a drop in the upper bound, as can be seen in figure~\ref{fig:OSgPM1}.
Instead of the plateau that we found in the supergraviton case (recall figure \ref{fig:CSgPM1}), here the bound on $\widetilde{\lambda}_{\phi}^2$ falls off continuously as the cutoff is increased. The string theory maximal cutoff $M^2=2m_\phi^2$ is not singled out in any way, nor is any other point along the boundary, and we see that the Type I amplitude \eqref{eq:MopenST} sits close to but below the boundary. It is unlikely that it will reach the boundary in the limit of infinite wavepackets and null constraints.

\begin{figure}[!ht]
     \centering
     \begin{subfigure}[b]{0.495\textwidth}
         \centering
         \includegraphics[width=\textwidth]{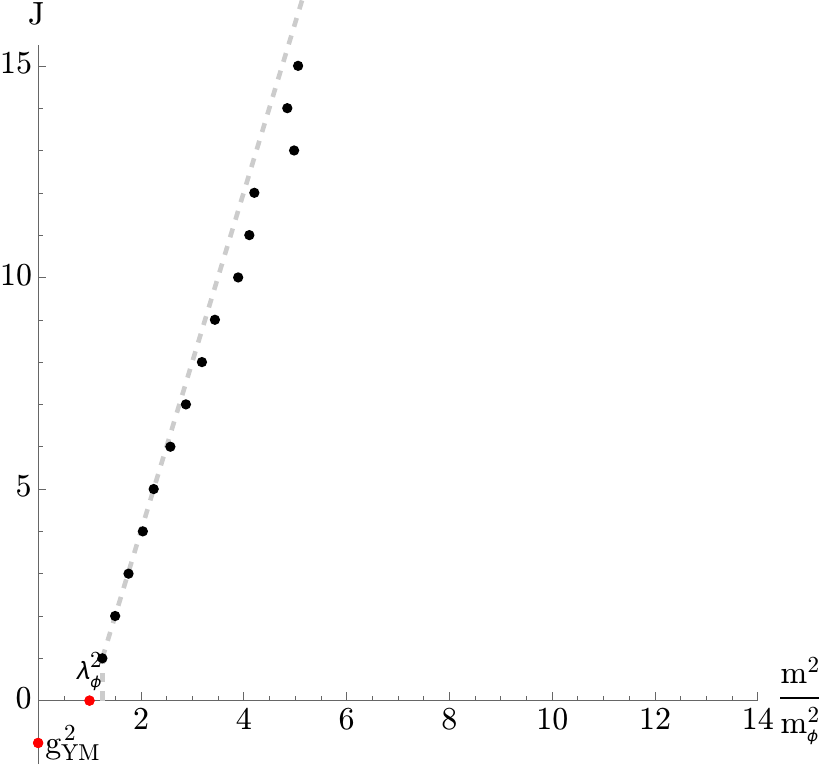}
         \caption{$M^2=1.25m_{\phi}^2$}
         \label{fig:OSspec800}
     \end{subfigure}
     \hfill
     \begin{subfigure}[b]{0.495\textwidth}
         \centering
         \includegraphics[width=\textwidth]{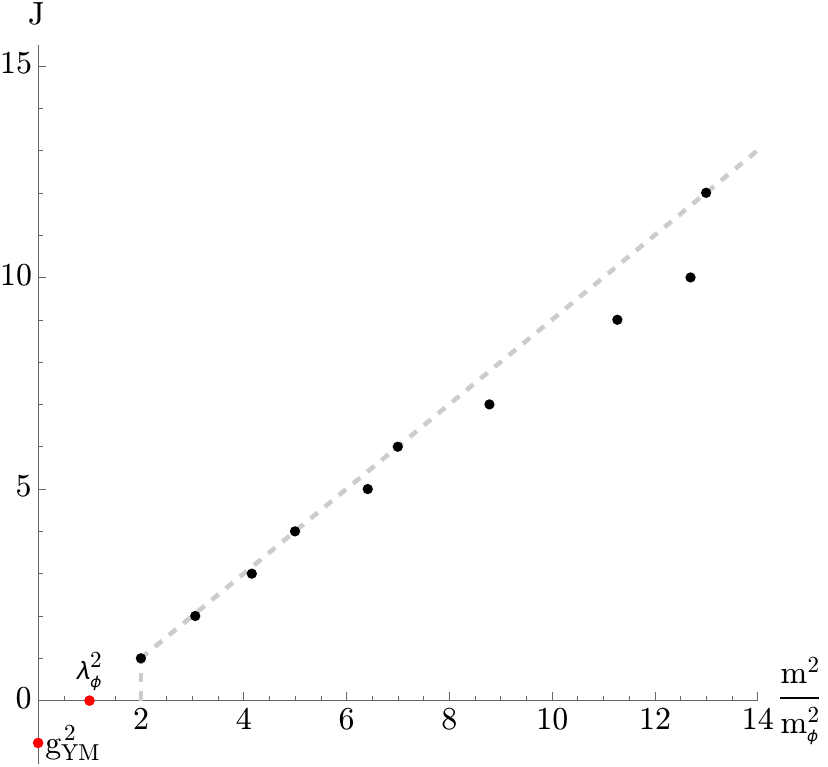}
         \caption{$M^2=2m_{\phi}^2$}
         \label{fig:OSspec500}
     \end{subfigure}
     \hfill
     \begin{subfigure}[b]{0.495\textwidth}
         \centering
         \includegraphics[width=\textwidth]{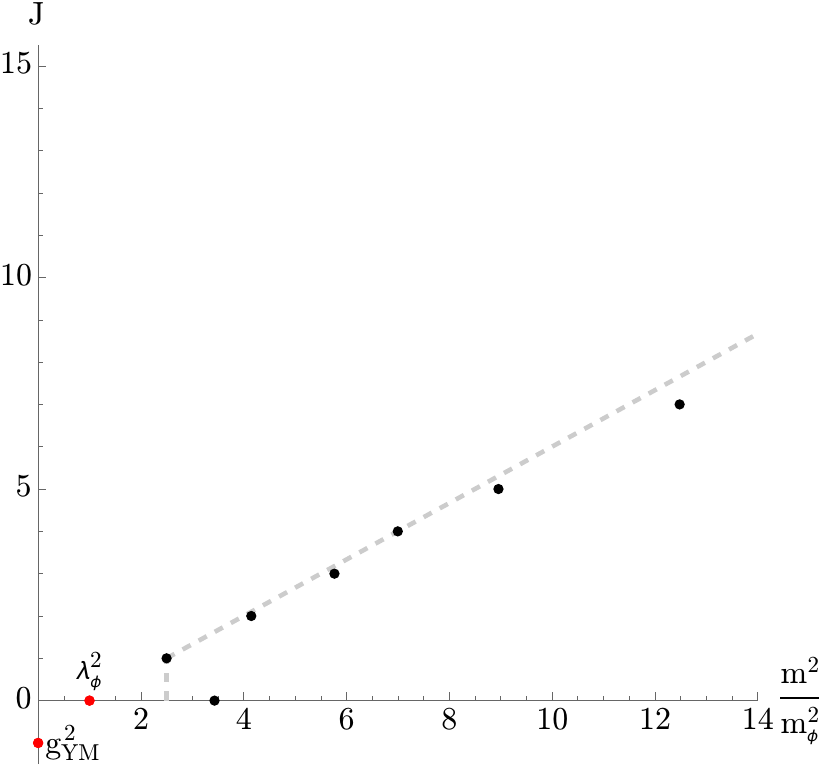}
         \caption{$M^2=2.5m_{\phi}^2$}
         \label{fig:OSspec400}
     \end{subfigure}
     \hfill
     \begin{subfigure}[b]{0.495\textwidth}
         \centering
         \includegraphics[width=\textwidth]{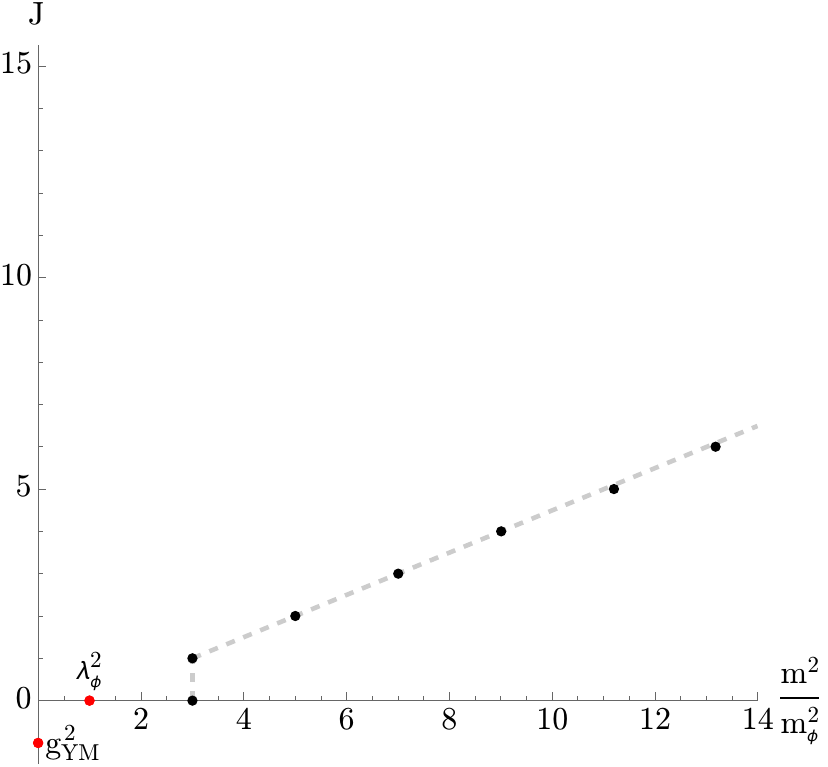}
         \caption{$M^2=3m_{\phi}^2$}
         \label{fig:OSspec333}
     \end{subfigure}
        \caption{Spectrum output from \texttt{SDPB} along the bound of figure \ref{fig:OSgPM1}. The red dots mark the gluon (which has effective spin $J_{\text{eff}}=-1$) and the enforced massive scalar. Maximal spectral assumptions that leave the bound fixed were imposed ---the resulting spin-dependent cutoff is shown by the grey dashed lines. Positivity is enforced for $m^2$ to the right of it. For all cutoffs we find that the states lie roughly along a line whose slope is dictated by the cutoff. Some double states as well as outliers are found, but we expect these to be due to the restriction to a finite number of wavepackets and null constraints.
        The plots \subref{fig:OSspec800} and \subref{fig:OSspec500} were generated with $k_{\text{max}}=13$ and $n_{\text{max}}=8$.
        Plots \subref{fig:OSspec400} and \subref{fig:OSspec333} were generated with $k_{\text{max}}=13$ and $n_{\text{max}}=6$.
        }
        \label{fig:OSspectra}
\end{figure}

If we look at the \texttt{SDPB} spectrum for various cutoffs (and make further spectral assumptions that do not change the bound but clean up the output spectrum), we find the family of extremal amplitudes shown in figure~\ref{fig:OSspectra}.
We see that the amplitudes that maximize the scalar coupling $\widetilde \lambda_\phi^2$ all have a linear Regge trajectory passing through the enforced scalar resonance at mass $m_\phi^2$ and a spin-one state appearing at the cutoff $M^2$. The cutoff thus controls the slope of the Regge trajectory, which is given by $m_{\phi}^2/(M^2-m_{\phi}^2)$. The scarcity of states below the linear trajectories of figure~\ref{fig:OSspectra} makes us conjecture that, just like those from section~\ref{sec:numAnalysis}, these extremal solutions too contain \textit{a single Regge trajectory} (besides the gluon pole). We interpret the numerical deviations from this spectrum as artifacts of the various truncations that went into the numerics. The two additional scalars in figures \ref{fig:OSspec400} and \ref{fig:OSspec333} are also spurious and can be removed by making additional assumptions on the spectrum that do not change the scalar coupling bound.

The main distinction between this putative family of extremal solutions for supergluon scattering and the one for supergravitons identified in section~\ref{sec:numAnalysis} is the effect of their free parameter. In the case of supergravitons, the free parameter controls the intercept of the trajectory, whereas in the supergluon case, it controls the slope. When the slope tends to infinity, the trajectory becomes vertical and we recover a tower-of-states amplitudes like~\eqref{eq:MtowerOpen}. On the other end, when the slope tends to zero, states with $J>0$ are pushed to infinity and the coupling of the scalar vanishes, leaving the pure SYM gluon pole alone. Understanding the precise mechanism by which all these single-trajectory-amplitudes are ruled in would be important progress in our understanding of positivity bounds.

\section{Outlook}

\label{sec:Outlook}

We have investigated the tree-level completions of  maximal 10D supergravity
that satisfy the minimal set of physical assumptions spelled out in the introduction. We have carved out theory space by systematically imposing the constraints
encoded in $2\to 2$ supergraviton scattering.
Our numerical bounds are rigorous, and they seem to have more or less converged to the optimal ones expected in the limit of infinitely many null constraints. Type II string theory lives safely within the allowed region in parameter space. Our most remarkable finding is an intriguing extremal solution of the dual bootstrap problem, which appears to consist of  a single, linear Regge trajectory with the same slope as  string theory. 

There are some obvious questions. First, 
what is the status of our extremal solution? The most conservative scenario is that an amplitude with a single linear Regge trajectory is only an approximate solution to the bootstrap constraints, perhaps in the sense that daughter trajectories can be systematically pushed to higher and higher energies but the limiting amplitude violates one of our hypotheses (e.g.~the Regge behavior). Indeed there is a rigorous no-go theorem in this sense~\cite{Eckner:2024pqt}. We should however be open minded.
Perhaps a fully consistent amplitude (at the level of $2\to 2$ supergraviton scattering) does exist, but it is exotic in some way, such as exhibiting Regge cuts, whose absence is one of the assumptions of~\cite{Eckner:2024pqt}.  
Of course, 
we have only imposed the simplest set of constraints;
full consistency requires considering the set of all amplitudes with arbitrary external states. 
The default conjecture remains that Type II string theory is the only {\it fully consistent} tree-level extension. Can we corner string theory by imposing the constraints encoded in a finite set of $2\to 2$ amplitudes?

The natural next step of our program is to consider the mixed system of $2\to 2$ amplitudes where the supergraviton and the first massive multiplet
can appear as external legs~\cite{inprogress}.
Besides its obvious relevance for the current problem, this will also serve as a very useful benchmark for 
the constraining power of
positivity bounds more generally, as superymmetry will vastly simplify the 
technical analysis compared to e.g.~the case of multiple amplitudes for the large $N$ meson bootstrap \cite{rhos}.

We have also studied the space of 
tree-level UV completions of 10D super Yang-Mills theory, focussing on
those that share some basic features with
open string theory. Again string theory (Type I in this case) strictly satisfies  our bounds. However here there
seems to be more flexibility. We find a whole one-parameter family of extremal solutions,
which appear to contain a single linear Regge
trajectory of varying slope. 
Similar 
questions arise as in the supergraviton case.
In particular we expect the system 
of amplitudes
where we allow the lighest massive supermultiplet as an external state to vastly
reduce the space of possible solutions. 
It would also be interesting to 
generalize our setup to the mixed SYM/gravity problem,  with the goal of cornering the heterotic string, where the graviton and the gluon are both in the closed string sector.

\section*{Acknowledgments}
We are grateful to  Vasiliy Dommes, Rajeev Erramilli, Pietro Ferrero, Felipe Figueroa, Yue-Zhou Li, Dalimil Maz\'{a}\v{c}, Matthew Mitchell, Martin Ro\v{c}ek, David Simmons-Duffin, Peter van Nieuwenhuizen, Piotr Tourkine and Sasha Zhiboedov for interesting discussion and comments.
We thank Henriette Elvang and Justin Berman for sharing a draft of~\cite{Berman:2024wyt} prior to publication.
This work was supported in part 
by the National Science Foundation under Grant NSF PHY-2210533 and by the  Simons Foundation under Grant 681267 (Simons Investigator Award).

The authors would like to thank Stony Brook Research Computing and Cyberinfrastructure and the Institute for Advanced Computational Science at Stony Brook University for access to the high-performance SeaWulf computing system, which was made possible by \$1.85M in grants from the National Science Foundation (awards 1531492 and 2215987) and matching funds from the Empire State Development’s Division of Science, Technology and Innovation (NYSTAR) program (contract C210148). 

\appendix

\section{Details on numerics}\label{app:numerics}
In this appendix we discuss how the bootstrap equations are transformed into a numerical problem for \texttt{SDPB} and list the different parameter regimes used in the numerics of this paper.
\subsection{Setting up SDPB}\label{app:sdpb}
 In general we are trying to find a vector $\Vec \alpha$ that maximizes the objective $\Vec \alpha \cdot \Vec v_{\text{obj}}$ while keeping the constraints
\begin{equation}
    \Vec \alpha \cdot \Vec v_{\text{norm}} = 1, \quad \Vec \alpha \cdot \Vec v_{\text{HE}}(m^2,J)\geq 0.
\end{equation}
This should hold for all $m\geq M$ and for supergravitons all even $J$, for supergluons even and odd $J$. In practice we deal with these infinite parameter ranges by splitting the parameter space into various regimes. Within each we choose appropriate discretizations and approximations. We explain each regime in detail in appendix \ref{app:parameters}. 

\texttt{SDPB} imposes positivity on polynomial inputs for $x\geq 0$. We have the freedom to, for every $J$, multiply $\Vec v_{\text{HE}}$ by a positive quantity. We can use this to make $\Vec v_{\text{HE}}$ polynomial. Generically this is done by multiplying $\Vec v_{\text{HE}}$ by its highest negative power of the variable in question (e.g. $m^2$ or $b$). Making one of the following substitutions then brings $\Vec v_{\text{HE}}$ into the necessary form. To enforce positivity in a one-sided open ended range, e.g. for $M^2\leq m^2 < \infty$, take $m^2 \rightarrow M^2 + x$. To enforce positivity in a window $m^2_{\text{min}}\leq m^2\leq m^2_{\text{max}}$ one should take $m^2 \rightarrow (m^2_{\text{min}} + m^2_{\text{max}}x)/(x+1)$ and follow by the necessary multiplication of powers of $(x+1)$. In cases where variables cannot be taken to be polynomial (or approximate them to be) one can instead evaluate $\Vec v_{\text{HE}}$ at discrete points.

\subsection{Parameter Regimes}\label{app:parameters}
\begin{figure}[ht]
\centering
\includegraphics[width=0.9\textwidth]{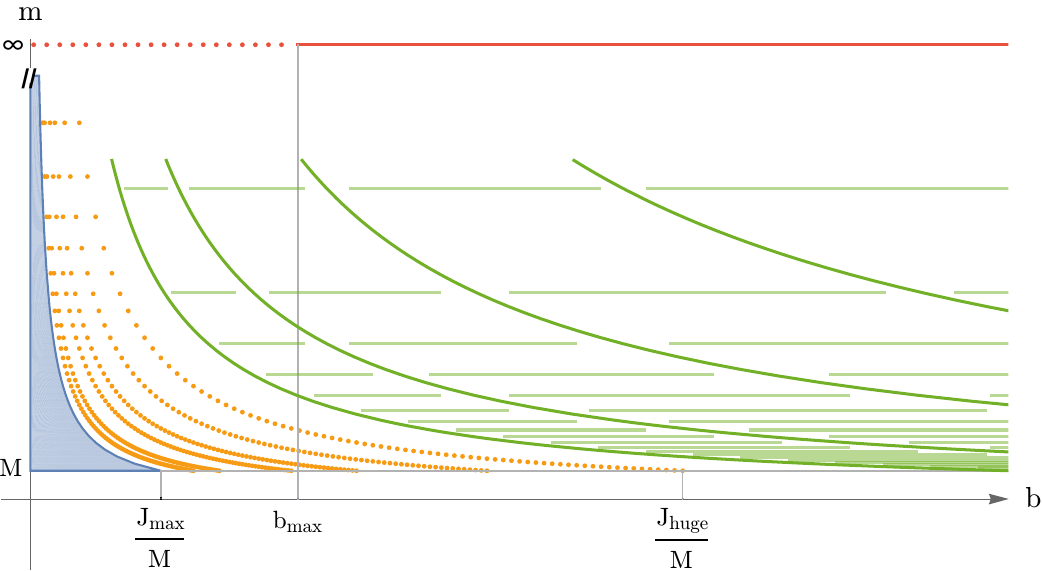}
\caption{Schematic of the various different parameter regimes in $b=2J/m$ vs.\ $m$ space. Continuous lines denote that the variable is treated polynomially, dots correspond to discrete points being probed. The blue region corresponds to the ``small $J$'' regime. The orange points correspond to the ``large $J$'' regime. The green lines correspond to the ``huge $J$, small $m$'' regime, with the darker constant $J$ ones being used for both supergravitons and supergluons, and the paler constant $m$ ones only being used in the supergraviton case. The red points correspond to the ``small $b$'' regime, the continuous red line corresponds to the ``large $b$'' regime.}
\label{fig:NumSchematic}
\end{figure}

 We explain each regime in detail below and have summarized them in table \ref{tab:parregime} and in the schematic in figure \ref{fig:NumSchematic}.
 
For bounds without spectral assumptions, i.e.\ those in sections \ref{sec:CSeftbounds} and \ref{sec:OSeftbounds} the parameter regimes used closely mirror those worked out in \cite{Caron-Huot:2021rmr}. We find that including spectral assumptions, i.e.\ including explicit resonances below the cutoff, requires an additional parameter regime (see ``very large $J$, small $m$'').

In general, increasing the number of null constraints and wavepackets requires including higher $J$ (or alternatively higher $b=2J/m$) and higher $m$ (to a lesser degree) to ensure convergence. For very small $k_{\text{max}}$ and $n_{\text{max}}$ only imposing positivity in the ``small $J$'' regime is sufficient.

\paragraph{small $\boldsymbol J$:}

We make no approximations at small $J$. Multiplying $\Vec v_{\text{HE}}$ by suitable powers of $m^2$ (this is allowed since $m^2\geq M^2\geq 1$) gives a vector of polynomials in $m^2$. \texttt{SDPB} then enforces positivity for the full range of $m^2$ and for all $J\leq J_{\text{max}}$ ($J$ even for supergravitons, $J$ even and odd for supergluons). In practice we are restricted to $J_{\text{max}}=60$. This regime is represented by the blue shaded region in figure \ref{fig:NumSchematic}.

\paragraph{large $\boldsymbol J$:}

Due to the smearing of the anti-subtracted sum rule the maximal degree of the polynomial in $m$ (for large enough $J$) of $\Vec v_{\text{HE}}$ is $2J+2$. For $J>J_{\text{max}}$ the polynomials become too large to work with as is. Instead we evaluate $\Vec v_{\text{HE}}$ at discrete points in both $J$ and $m^2$. The mesh we use is chosen to be densest for smaller $J$ and close to the cutoff. It then slowly widens as the two variables are increased. This regime is imposed for spins $J_{\text{max}}< J \leq J_{\text{huge}}$ where for us $J_{\text{huge}} = 5000$.\footnote{Larger $J_{\text{huge}}$ of up to 16k were tried but other than drastically increasing runtime did not have any effect on the numerics. EFT coefficient bounds all converge significantly earlier and bounds with spectral assumptions required the huge $J$ regime to converge at all.} The set of orange points in figure \ref{fig:NumSchematic} represents this regime.

\paragraph{huge $\boldsymbol J$, small $\boldsymbol m$:} Imposing the positivity constraint exactly for spins above $J_{\text{huge}}$ quickly becomes impossible numerically. The runtime increases logarithmically with the maximal spin. Fortunately when no spectral assumptions are made the numerics converge quickly and well before this limit. The situation changes though when the scalar resonance is imposed and the cutoff is increased. Roughly speaking, at fixed $n_{\text{max}}$ impact parameters up to a finite fixed $b_{\text{conv}}$ are probed. Increasing the cutoff requires including more and more spins to reach $b_{\text{conv}}$. For small $m$ and very large $J> J_{\text{huge}}$ we find that the null constraints and forward limit sum rules dominate. In this regime we can therefore approximate $\Vec v_{\text{HE}}$ by setting the integrated sum rules to zero and keeping only the forward limit ones, for example \eqref{eq:CSlowJvec} becomes:
\begin{equation}
    \Vec v_{\text{HE}} \approx \left(0,\ldots, 0, 0, \mathcal X_{1,0}(m^2,J), \cdots \right)\,.
\end{equation}
As a reminder, this approximation at large $J$ only holds for relatively small $m^2$. Within this regime though we can now discretize in $J$ and take $m^2$ polynomial within a window $M\leq m \leq m_{\text{max}}$, see bolder green (curved) lines in figure \ref{fig:NumSchematic}. This is the best we can do in the supergluon case. For supergravitons, as all spins are restricted to be even, the oscillations $(-1)^J$ in the null constraints do not play a role. $J$ therefore appears polynomially in $\Vec v_{\text{HE}}$. We can then further make the approximation that $J$ is continuous and impose the positivity constraint for discrete (small) $m^2$ with polynomial $J$. This is imposed between the discrete $J$ which were polynomial in $m^2$ and allows for us to impose the constraint in a grid-like fashion. The polynomial in $J$ segments are represented by the paler green (horizontal) lines in figure \ref{fig:NumSchematic}.

\paragraph{small $\boldsymbol b$:}
We also impose the constraints in the limit $J,m \rightarrow \infty$ with $b=\frac{2J}{m}$ fixed. The Gegenbauer polynomials \eqref{eq:Gegenbauer} are well understood in this limit and become Bessel functions, see Appendix A of \cite{Caron-Huot:2021rmr} for details. In this limit the sum rules with the lowest number of subtractions dominate, i.e.\ the smeared sum rules, and the rest can be set to zero. $\Vec v_{\text{HE}}$ is oscillatory in $b$. For small $b$ we therefore again impose the constraint at discrete points within the range $0< b\leq b_{\text{max}}$ and refine the mesh where needed. 

\paragraph{large $\boldsymbol b$:}
At large $b \geq b_{\text{max}}$ it was explained in \cite{Caron-Huot:2021rmr} that one could approximate and rewrite the constraint into a vector of $2\times 2$ matrices, polynomial in $b$. Details can be found in the original reference. Since this approximation is a stronger constraint than the original constraint the bound can become weaker. For this not to affect the result a large enough $b_{\text{max}}$ must be chosen. For most numerics $b_{\text{max}}=80$ is sufficient. For precise bounds at four or more significant figures this needed to be increased to $b_{\text{max}}=200$.

\paragraph{A note on discretization:} In both the large $J$ and small $b$ regime some sort of discretization was used. Since positivity is only being enforced at points rather than a continuum it can happen that the action of $\Vec \alpha \cdot \Vec v_{\text{HE}}$ becomes negative between two points. As was explained in \cite{Caron-Huot:2021rmr} refinements of the mesh will be needed in this case. This was done by examining the functional output by \texttt{SDPB}, finding areas of negativity, and rerunning the numerics having added extra intermediate points to cover these areas. This process was repeated until it only resulted in insignificant changes to the resulting bound.

\begin{table}
\centering
\begin{tabular}{|c|l|}
\hline
\textbf{parameter range}                                 & \multicolumn{1}{c|}{\textbf{implementation of constraint}}                          \\ \hline
$0\leq J\leq J_{\text{max}}$                                   & \multicolumn{1}{l|}{all $J$, polynomial in $m^2$}               \\ \hline
$J_{\text{max}} < J \leq J_{\text{huge}}$ & \multicolumn{1}{l|}{discrete in $J$ and $m^2$}                  \\ \hline
$0<b=2J/m < b_{\text{max}}$                      & \multicolumn{1}{l|}{discrete in $b$}                            \\ \hline
$b\geq b_{\text{max}}$                                   & \multicolumn{1}{l|}{$2\times 2$ matrix approximation, polynomial in $b$} \\ \hline \hline
$J\geq J_{\text{huge}}$ &
  \multicolumn{1}{l|}{\begin{tabular}[c]{@{}l@{}}neglect smeared sum rules, discrete $J$, polynomial in $m^2$\\ \textit{additionally for supergravitons:}  polynomial in $J$, discrete $m^2$ \end{tabular}} \\ \hline
\end{tabular}
\caption{Summary of various approximations and discretizations used in numerics. The $J\geq J_{\text{huge}}$ regime is only needed when spectral assumptions (i.e. imposing the scalar at mass 1 and pushing the cutoff above 1) are made.}
\label{tab:parregime}
\end{table}

\bibliographystyle{ytphys}
\bibliography{references}

\end{document}